\documentclass[
reprint,
 aip,
 jcp,
amsmath,amssymb
]{revtex4-2}

\usepackage{graphicx}
\usepackage{subcaption}
\usepackage{dcolumn}
\usepackage{bm}
\usepackage{enumitem}
\usepackage[normalem]{ulem}
\usepackage{cancel}

\usepackage{lmodern}  
\usepackage{physics}
\usepackage{xcolor}

\begin{document}

\preprint{AIP/123-QED}

\title[Magnus expansions for two-level quantum dynamics]{Higher order Magnus expansion for driven two-level quantum dynamics}

\author{Chen Wei}
\author{Frank Großmann}%
 \email{frank.grossmann1@tu-dresden.de}
\affiliation{
Institut f\"ur Theoretische Physik, Technische Universit\"at Dresden, 01062 Dresden, Germany
}%


\date{\today}

\begin{abstract}
We investigate the Magnus expansion for a generic time-dependent two-level system
under single-axis driving. 
By virtue of the \(\mathfrak{su}(2)\) Lie algebra, the expansion is decomposed into a commutator-free form. To illustrate the usefulness of the gained expression, we then revisit the Landau-Zener-St\"uckelberg-Majorana model, with a focus on non-adiabatic transitions as well as the Stokes phase. In addition, the semiclassical Rabi model is systematically treated by determining the Floquet quasienergy up to different orders. 
We demonstrate how to employ suitable picture transformations as well as on how to enforce the symmetry of the underlying model in order to guarantee convergence of the expansion as well as to achieve satisfactory agreement with the exact results. For both models that we studied it turns out that a third order approximation yields results that are in next to perfect agreement with exact analytical ones. 
Surprisingly, in the case of the semiclassical Rabi model, even the second order Magnus approximation in the adiabatic picture produces almost exact results  for a large parameter range.
\end{abstract}


\maketitle
\section{Introduction}\label{sec:intro}
In this work we investigate the Magnus 
expansion\cite{Magnus1954} (ME) with a 
focus on two-level quantum dynamics under 
single-axis driving, modeled by the Hamilton 
operator
\begin{equation}\label{eq:phys_hamiltonian}
 H(t)=\frac{\Delta}{2}\sigma_z+\frac{f(t)}{2}\sigma_x,
\end{equation}
with Pauli matrices $\sigma_z,\sigma_x$ and where \(\Delta\) is a (positive) constant while \(f(t)\) may be arbitrarily time-dependent. Hamilton operators in the form of Eq.~\eqref{eq:phys_hamiltonian} have a long tradition starting in the heyday of quantum theory, with the works of Landau, Zener, St\"uckelberg and Majorana (LZSM) \cite{Landau1932, Zener1932,Stueckelberg1932,Majorana1932}, as well as Rosen and Zener (RZ) \cite{RosenZener1932}. In addition, the semiclassical Rabi model \cite{Rabi1936,Rabi1937,BS-shift1940,Braak_2016}  
with a linearly polarized drive
falls into this class. It is striking that although the quantum dynamics of LZSM and RZ models have long been known to be solvable analytically exactly, the 
semiclassical Rabi model has only been solved analytically without employing the rotating wave approximation in the early 21st century \cite{MaLi2007, XieHai2010}. Furthermore, alternative (as well as more insightful) derivations and interpretations of Landau-Zener transitions are published up to the present days \cite{Wittig2005,Vuhta2010,Ho2014,Glasbrenner2023,Glasbrenner2025,Sun2025} and the relevance of the above models in different fields of physics and chemistry cannot be overemphasized. Applications are ranging from molecular collisions \cite{TullyPreston1971,Cindro1986} and driven quantum tunneling \cite{CDT-PRL-1991,epl92,GLP92,Wang1994,Kayanuma1994,Grifoni1998,Kayanuma2008} 
to quantum state preparation \cite{Cao2000,Saito2006,Ribeiro2009}, to name but a few.

The interest in two-level quantum dynamics has persisted in recent times, driven by the increasing relevance of qubit dynamics and control \cite{epl92,VanDamme2017}. For the well-known exactly solvable cases mentioned above, non-elementary special functions are frequently encountered, making them impractical for applications and physical interpretation, however. In order to mitigate this problem, reverse engineering approaches have been developed\cite{GangopadhyayPRB2010, BarnesPRL2012}, enabling the construction of infinitely many analytically solvable driving functions \(f(t)\). However, it remains unclear how the general dynamical behavior is determined by external driving with, e.g., pulsed laser fields. For example, previous studies have observed that although the pulse shapes may be qualitatively similar, the system’s evolution will depend sensitively on their precise form in the case of non-resonance of the drive\cite{BermanPRA1998, ConoverPRA2011, RalfPRD2015}. 


Groundbreaking efforts to understand general aspects of two-state quantum dynamics have been made, particularly in the context of non-adiabatic transitions. Based on the work of  Davis and Pechukas\cite{DavisPechukas1976} and Berry\cite{Berry1990Adiabatic}, non-adiabatic transitions can be understood by analyzing the singularities of the Hamiltonian under analytic continuation into the complex-time plane. This approach yields the correct exponent for the LZ transition probability and the underlying ideas were taken up by the so-called exact WKB analysis\cite{MatsudaWKB2025}, which has been extensively developed and discussed in the past few years. While this approach is promising, its mathematical complexity and the challenges associated with generalizing to multi-level systems limit its practical appeal. As another commonly applied approach, the Dyson series\cite{FGbook} converges for quantum systems with a finite-dimensional Hilbert space at any fixed time. Nevertheless, issues such as wave-function normalization and the emergence of secular terms require special care
\cite{ RevModPhys.44.602}. More fundamentally, non-perturbative effects are often exponentially small in the coupling strength and cannot be captured by any finite-order or even infinite-order perturbative expansion. A prominent example of such behavior is the so-called Landau-Zener transition in the LZSM model.

In contrast to the two approaches discussed above, the ME for the time-evolution operator exhibits an elegant simplicity. By construction, it automatically preserves the unitarity of the time-evolution operator for a Hermitian Hamiltonian, and can therefore considerably simplify the calculation. Specializing to two-level systems, the ME has been successfully applied to the Rosen-Zener problem\cite{BLANES2009151}, Rashba 
spin-orbit coupling \cite{Lopez2013graphene}, Rabi models in the context of nuclear magnetic resonance
\cite{Haeberlen1968,Evans1968,Feldman1984} and under chirped pulses\cite{NalbachPRA2018} as well as to non-Hermitian two-level systems\cite{Begzjav2020103098}. Furthermore, in half-cycle pulse driven hydrogen atoms, different orders of ME have been used in different gauges to reach the same accuracy \cite{Klaiber2008}. As will be shown in the present paper, a low order ME approach serves as a quickly converging approximation for the LZSM problem, 
as well as for the semiclassical Rabi model thereby indicating its ability to describe non-perturbative effects.


Yet the ME has its own subtleties; even in the simplest case of a two-level system, it can diverge when the evolution time becomes large\cite{Pechukas1966Magnus,Salzmann1986CPL}. By now, the mathematical aspects of the convergence of the ME are well-studied\cite{Maricq1987, jakubovič1975linear,Blanes_1998, Casas_2007}. For physical applications, on the other hand, properties of convergence can be drastically different in different pictures of quantum mechanics, as noticed by several authors\cite{Salzman1986,Maricq1987, BLANES2009151}. The importance of convergence studies and the usefulness of picture transformations, however, have not yet been fully exploited in previous work. In the present contribution, we will make further progress in this direction. We show that it is constructive to study different parameter regimes of the Hamiltonian Eq.~\eqref{eq:phys_hamiltonian}, requiring only the qualitative features of  \(f(t)\) to be specified. In particular, we prove that when \(f(t)\) is a monotonic function, the ME in the adiabatic picture is ensured to converge. This also provides a quantitative characterization of the adiabatic theorem \cite{Born1928beweis} in the simple case of two-level systems. 




The article is structured as follows: In Sec.~\ref{sec:su2} we lay the foundation for the rest of the article by employing the \(\mathfrak{su}(2)\) algebra to reduce the ME for the two-level problem into commutator-free form, the detailed derivation of which can be found in Appendix \ref{sec:magnus_app}.  Sec.~\ref{sec:non-adiabatic} is then devoted to the reinvestigation of the non-adiabatic problem, emphasizing the importance of the adiabatic picture. Our investigations are exemplified by the LZSM problem. In Sec.~\ref{sec:Floquet} a novel analytic computational scheme for the Floquet quasienergy of a periodically driven two-level system is proposed, based on the ME and unitary transformations. We systematically apply the scheme to the semiclassical Rabi model. 
The ME is found to be a versatile tool that is even more broadly applicable 
as well as more powerful (quickly converging) than it might have been expected. We conclude with a discussion of the presented results and the future potential 
of the ME in Sec.~\ref{sec:discussion}. Two further Appendices \ref{sec:symmetry_app} and \ref{sec:crossing_app} deal with the symmetries of the problems discussed and with the different types of quasienergy crossings encountered, respectively.
\section{SU(2) group and Magnus expansion}\label{sec:su2}

The quantum mechanical time-evolution operator \(U(t,t_0)\) allows to propagate an arbitrary initial state from a fixed initial time \(t_0\) to the final time $t$ and has the initial condition \(U(t_0, t_0)=\mathbb{I}\),
with $\mathbb{I}$ denoting the identity operator. For notational convenience, we mostly drop the initial time argument and use the fact that \(U\) can be written in exponential form 
\begin{equation}\label{eq:magnus_operator}
    U(t)=\exp\left[-i\Omega(t)\right]
\end{equation}
for all times\cite{Magnus1954}.
\(\Omega(t)\) is referred to as the Magnus operator which, in general, is given in terms of an infinite series, to be explicitly detailed below for the case of dynamics under time-dependent Hamiltonians of the form of Eq.~\eqref{eq:phys_hamiltonian}. 

Since the Hamiltonian in Eq.~\eqref{eq:phys_hamiltonian} is traceless, at each point of time, the corresponding time-evolution operator \(U(t)\) belongs to the SU(2) group and admits the standard angle-axis parameterization\cite{Hall2015LieGroups}. I.e., there exists a real-valued function \(\theta(t)\) and a unit vector \(\vec{n}(t)\) such that
\begin{align}
\stepcounter{equation}
    U(t)&=\exp \left[-i\theta(t) \vec n(t) \cdot \vec \sigma\right] \label{eq:U_angel_axis}\\
    &=\cos\theta~\sigma_0-i\sin\theta~\vec{n}\cdot\vec\sigma \label{eq:su2_exp}
\end{align}
where \(\vec{\sigma}=(\sigma_x, \sigma_y, \sigma_z)\) is a vector whose elements are the Pauli matrices and \(\sigma_0\) is the matrix representing the \(2\times 2\) identity operator. Comparing Eq.~\eqref{eq:U_angel_axis} with Eq.~\eqref{eq:magnus_operator}, the Magnus operator can be identified as
\begin{equation}\label{eq:Omega_angle_axis}
    \Omega(t)=\theta(t) \vec n(t) \cdot \vec \sigma.
\end{equation}
and it belongs to the \(\mathfrak{su}(2)\) algebra,  i.e., it can be represented by the set of traceless Hermitian \(2\times 2\) matrices. The initial condition \(U(t_0)=\sigma_0\) is then  accounted for by \(\theta(t_0)=0\). 

The raising and lowering operators
\(\sigma_{\pm}:=\frac{1}{2}(\sigma_x\pm i\sigma_y)\), 
together with $\sigma_z$, fulfill the commutation relations
\begin{equation}\label{eq:su2_algebra}
    [\sigma_z, \sigma_\pm]=\pm 2\sigma_\pm, \ [\sigma_+, \sigma_-]=\sigma_z
\end{equation}
and form a complete basis for 
\(\mathfrak{su}(2)\). Thus \(\Omega(t)\) can be decomposed as 
\begin{equation}\label{eq:Omega_decomp}
    \Omega(t)= A(t)\sigma_+ +A^*(t)\sigma_- + C(t)\sigma_z,
\end{equation}
where the coefficient \(A(t)\) is complex in general, while \(C(t)\) has to be real-valued. Comparing Eq.~\eqref{eq:Omega_angle_axis} with Eq.~\eqref{eq:Omega_decomp}, the angle and axis can be expressed by these coefficients as
\begin{eqnarray}\label{eq:angle_axis_AC_1}
    \theta=\sqrt{|A|^2+|C|^2}, 
    \\
    \label{eq:angle_axis_AC_2}
    \vec n =\frac{1}{\theta}(\text{Re}(A), 
    -\text{Im}(A), C).
\end{eqnarray}
The quantities in Eqs.~(\ref{eq:angle_axis_AC_1},\ref{eq:angle_axis_AC_2}) will be the main objects of the calculations in this work. 

As reviewed in \cite{Tannor2007}, the Magnus expansion expresses \(\Omega(t)\) as an infinite series built from time integrals of the Hamiltonian \(H(t)\) and its nested multi-time commutators. Observing that the \(\mathfrak{su}(2)\) decomposition in Eq.~\eqref{eq:Omega_decomp} can be performed to each order of the ME, in Appendix~\ref{sec:magnus_app}, we derive an expansion directly for \(A(t)\) and \(C(t)\), reading
\begin{equation}\label{eq:AC}
    A(t)=\sum_{n=1}^{\infty} A_n(t),\quad 
    C(t)=\sum_{n=1}^{\infty} C_n(t),
\end{equation}
where \(A_n(t)\) and \(C_n(t)\) are scalar functions that can be viewed as decomposition coefficients of the n-th order ME and can be determined recursively. Later-on, we refer to them as the Magnus coefficients. The Magnus approximation is then achieved by truncating the summations in Eq.~\eqref{eq:AC} at certain orders.  


Up to this point, our discussion is valid for the most general two-level Hermitian Hamiltonian. The specific structure of the Hamiltonian may, however, provide further simplifications. While a Hamiltonian in the form of Eq.~\eqref{eq:phys_hamiltonian} admits 
for a direct physical interpretation, for reasons that will become clear later, we will switch to a Hamiltonian of the form
\begin{equation}\label{eq:hamiltonian_magnus}
    H(t)=v(t)\sigma_+ + v^*(t)\sigma_-,
\end{equation}
where \(v(t)\) is a complex scalar function, which can be expressed by \(f(t)\) and \(\Delta\) in different ways for different pictures of quantum mechanics, see for example Eq.~\eqref{eq:h_adiabatic} and Table~\ref{tab:table1} further below. 
For a Hamiltonian in the above special form, by using Eq.~\eqref{eq:ac_initial} and Eqs.~\eqref{eq:recursion_a}-\eqref{eq:recursion_c}, it can be recursively verified that
\begin{equation}\label{eq:vanishing_AC}
    A_{2n}(t)=C_{2n-1}(t)=0, \quad \forall n\geq1,
\end{equation}
i.e., all even indexed $A$-coefficients and all odd-indexed $C$-coefficients vanish, considerably simplifying our efforts. 
Defining \(v_1:=v(t_1),v_2:=v(t_2)\dots\), the three leading non-zero coefficients read

\begin{align}
A_1(t)&=\int_{t_0}^t{\rm d}t_1 v_1, \label{eq:A1}
\\
C_2(t)&=\int_{t_0}^t {\rm d}t_1\int_{t_0}^{t_1}\!{\rm d}t_2\;\text{Im}[v_1v^*_2],  \label{eq:C2}
\\
A_3(t) &=\frac{2i}{3}\int_{t_0}^t {\rm d}t_1\int_{t_0}^{t_1} {\rm d}t_2\int_{t_0}^{t_2} {\rm d}t_3
\notag
\\
&\qquad\Big[ \text{Im}(v_2 v_3^*)v_1 + \text{Im}(v_2 v_1^*)v_3 \Big].  \label{eq:A3}
\end{align}
Commutator free equations for the coefficients similar to the ones above have been found before directly from the ME \cite{Begzjav2020103098},   while in Appendix~\ref{sec:magnus_app} we employ the
\(\mathfrak{su}(2)\) decomposition to provide a systematic treatment. For algorithmic evaluation, the recursion relations of \(A_n(t)\) and \(C_n(t)\) derived there are much more efficient than the multi-variate integration form in Eq.~\eqref{eq:A1}-\eqref{eq:A3}. 

 We would like to point out that our formalism is based on a methodology that is essentially different from so-called commutator-free integrators~\cite{BLANES20061519, Alvermann2011} in numerical applications, where the evaluation of commutators is avoided by using the Baker–Campbell–Hausdorff formula and suitability choosing the quadrature rule. Furthermore, the idea of using ladder operators to achieve simplifications can also be found, for example, in~\cite{Rau2005} on the Wei-Norman method 
that expresses the time-evolution operator of a driven quantum system in terms of a 
product of exponential operators.

We now turn to the final aspect of the ME for two-level systems, namely the convergence criterion. According to Maricq\cite{Maricq1987}, a sufficient convergence condition is given by
\begin{equation}\label{eq:magnus_criteria_tls}
    \int_{t_0}^t E(s){\rm d}s < \pi,
\end{equation}
where \(E(t)\) is the positive instantaneous eigenvalue of the two-level traceless Hamiltonian. In the case 
of Eq.~\eqref{eq:hamiltonian_magnus}, we have
\begin{equation}
    E(t)=|v(t)|.
\end{equation}
We note that \(E(t)\) can be regarded as the operator norm of  a \(2\times 2\) traceless Hermitian matrix, and therefore Eq.~\eqref{eq:magnus_criteria_tls} is a special case of Casas' \(\pi\)-criterion\cite{Casas_2007} for general bounded Hamiltonians. The different performance of the ME in different pictures of quantum mechanics can be traced back to the fact that the norm of the Hamiltonian may differ drastically by switching to another picture as will be seen in Section \ref{sec:Floquet}.
\section{Non-adiabatic transition problem}\label{sec:non-adiabatic}
The (quantum) adiabatic theorem states that a quantum system initially prepared in an instantaneous eigenstate of an infinitely slowly varying Hamiltonian remains in the corresponding instantaneous eigenstate throughout the evolution, up to a phase factor \cite{Born1928beweis,Tannor2007}. The non-adiabatic transition problem deals with the determination of transition probabilities between instantaneous eigenstates of a time-dependent Hamiltonian when the adiabatic condition is violated. 

In the present section, we consider a two-level Hamiltonian with instantaneous eigenstates \(\ket{\phi_\pm (t)}\) and the system shall be prepared in one of those states in the asymptotic limit \(t\to t_0=-\infty\),
i.e.,
\begin{equation}\label{eq:a_init}
    \lim_{t\to -\infty}|\langle{\psi(t)}\ket{\phi_+(t)}|=1.
\end{equation}
The central task then is to find the final transition probability to the other instantaneous eigenstate, i.e., we want to determine 
\begin{equation}\label{eq:a_P}
    P(+\infty)=\lim_{t\to +\infty}|\langle{\psi(t)}\ket{\phi_-(t)}|^2.
\end{equation}
This problem can be most cleanly addressed in the adiabatic (interaction) picture\cite{Garrido1964, BerryAdiabaticPicture1987, BLANES2009151}, as reviewed below. We note that the 
ME in the adiabatic picture has been paid special attention to in previous works, such as\cite{Klarsfeld1992, BLANES2009151}. Our main target in this section is to point out that the convergence condition Eq.~\eqref{eq:magnus_criteria_tls}, which has this far not been in the center of interest, holds for a large class of drivings in Eq.~\eqref{eq:phys_hamiltonian}. This statement is then corroborated by applications of the formalism established in Sec.~\ref{sec:su2} to the LZSM model in the present section and to the semiclassical Rabi model in Sec.~\ref{sec:Floquet}.

\subsection{Adiabatic picture}
To start with, one defines the dynamical phase
\begin{equation}
\label{eq:dynamical_phase}
    \varphi(t): ={\int_{t_0'}^t} E(\tau){\rm d}\tau =\frac{1}{2}\int _{t_0'}^t \sqrt{\Delta^2+f^2(\tau)}{\rm d}\tau,
\end{equation}
where \(E(t)\) is the positive instantaneous eigenvalue of the Hamiltonian of 
Eq.~\eqref{eq:phys_hamiltonian} and the lower limit of integration $t_0'$ can be chosen arbitrarily. 

The adiabatic picture then follows by expanding the wave function in 
instantaneous eigenstates, according to
\begin{align}
    |\psi(t)\rangle&=\alpha(t) e^{-i\varphi(t)}|\phi_+(t)\rangle+\beta(t)e^{+i\varphi(t)}|\phi_-(t)\rangle\label{eq:bogoliubov_expansion}
    \\
&=\Psi(t)\Phi(t)\ket{\psi_a(t)}=:U_0(t)\ket{\psi_a(t)},\label{eq:psi_a}
\end{align}
where
\begin{align}
\Psi(t)&=\big[|\phi_+(t)\rangle, |\phi_-(t)\rangle\big], 
\\
\Phi(t)&=\exp\left[-i\varphi(t)\sigma_z\right],
\label{eq:Phi}
\end{align}
and \(\ket{\psi_a(t)}=[\alpha(t), \beta(t)]^T\) is denoting the wave function in the adiabatic picture. Comparing Eqs.~\eqref{eq:bogoliubov_expansion} and Eq.~\eqref{eq:a_init}, its initial condition  follows to be
\begin{equation}
    |\alpha(t_0)|=1, \quad |\beta(t_0)|=0
\end{equation}
and the non-adiabatic transition probability is given by
\begin{equation}
P(t)=|\beta(t)|^2.
\end{equation}

On the other hand, Eq.~\eqref{eq:psi_a} is equivalent to a unitary transformation under the evolution operator
\begin{align}
\label{eq:U_adiabatic_pic}
U(t)&=U_0(t)U_a(t)U_0^\dagger(t_0)
\end{align}
where 
\begin{equation}\label{eq:Ua-matrix}
U_a(t)=\begin{bmatrix}
    \alpha(t) & -\beta^*(t)\\
    \beta(t) & \alpha^*(t)
\end{bmatrix}
\end{equation}
denotes the time-evolution operator in the adiabatic picture. The presence of \(U_0^\dagger(t_0)\) in Eq.~\eqref{eq:U_adiabatic_pic} ensures the initial condition \(U(t_0)=U_a(t_0)=\mathbb{I}\).

Now we move on to determine the the Hamiltonian in the adiabatic picture, such that 
\begin{align}
i\dot U_a(t)=H_a(t)U_a(t).
\end{align} 
According to the Schrödinger equation for \(U(t)\) and the definition of \(U_0(t)\), a well as by using \(H(t)\Psi(t)=\Psi(t)[E(t)\sigma_z]\), one finds
\begin{equation}\label{eq:H_a_general}
    H_a(t)=-i\Phi^\dagger(t)\Psi^\dagger(t)\dot\Psi(t)\Phi(t).
\end{equation}
To proceed, one has to make use of the explicit expressions of instantaneous eigenstates \(\ket{\phi_\pm(t)}\) of the  
Hamiltonian in Eq.~\eqref{eq:phys_hamiltonian}, which can be chosen up to a time-dependent phase factor. In this work, we choose the real-gauge~\cite{Berry1990Adiabatic,Stenholm96},
leading to
\begin{align}
    &{|\phi_+(t)\rangle=\begin{bmatrix}
    \cos{[\chi(t)/2]} \\
    \sin{[\chi(t)/2]}
\end{bmatrix},}
\\ 
&|\phi_-(t)\rangle=\begin{bmatrix}
    -\sin{[\chi(t)/2]} \\
    \cos{[\chi(t)/2]}
\end{bmatrix},
\end{align}
where
\begin{equation}\label{eq:chi_angle}
     \chi(t) :=\arctan\left[{f(t)}/{\Delta}\right].
\end{equation}

It is clear from Eq.~\eqref{eq:chi_angle} that  \(\chi(t)\) does not increase beyond \(2\pi\) whenever \(f(t)\) evolves continuously, in contrast to the double axis driving considered in \cite{YingPRR2020}. In our case, $\ket{\phi_\pm(t)}$ thus is real as well as continuous, and the Berry phase \cite{BerryPhase1984} vanishes. 
Furthermore, by taking the derivative we have
\begin{equation}\label{eq:psi_dot}
\langle \dot \phi_+(t)|\phi_+(t)\rangle=\langle \dot \phi_-(t)|\phi_-(t)\rangle=0,
\end{equation}
 and according to Eq.~\eqref{eq:H_a_general}, the matrix representation of the adiabatic Hamiltonian is given by
\begin{align}
    H_a(t)
    &=\frac{\dot \chi(t)}{2} \begin{bmatrix}
    0 & i {\rm e}^{2i\varphi(t)}\\
     -i{\rm e}^{-2i\varphi(t)} & 0
\end{bmatrix}.\label{eq:h_adiabatic}
\end{align}

 In addition, one finds that Eq.~\eqref{eq:h_adiabatic} is in accord with the structure of the Hamiltonian in Eq.~\eqref{eq:hamiltonian_magnus}
with the identification 
\begin{eqnarray}\label{eq:v_adiabatic}
        v(t)&=& i\frac{\dot \chi(t)}{2}  {\rm e}^{2i\varphi(t)},
        \\
        \label{eq:chi_adiabatic}
        \dot\chi(t)&=&\frac{\Delta\dot f(t)}{\Delta^2+f^2(t)}.
\end{eqnarray}

It is striking that the (positive) instantaneous energy of \(H_a(t)\), namely \(E_a(t)=|v(t)|=\frac{|\dot \chi(t)|}{2}\), is very different from that of the original Hamiltonian in Eq.~\eqref{eq:phys_hamiltonian}. Therefore, according to Eq.~\eqref{eq:magnus_criteria_tls}, one may expect that the convergence property of the Magnus expansion is drastically different among different pictures. Indeed, remarkably, in the time interval where 
\(f(t)\) is monotonic, the convergence condition Eq.~\eqref{eq:magnus_criteria_tls} is always satisfied in the adiabatic picture. To see this, we first assume that 
\(f(t)\) monotonically increases on \(t\in[t_0, t_1]\), and hence \(\dot f(t)>0\). Then according to Eq.~\eqref{eq:v_adiabatic}, $|\dot \chi(t)|$ can be replaced with $\dot \chi(t)$
\begin{equation}
   \notag
\end{equation}
and hence by using the definition in Eq.\ (\ref{eq:chi_angle}), we find 
\begin{equation}\label{eq:chi_bound}
    \int_{t_0}^{t_1}E_{a}(\tau){\rm d}\tau=\frac{1}{2}[\chi(t_1)-\chi(t_0)] \leq \frac{\pi}{2}<\pi.
\end{equation}
One may draw the same conclusion if  \(f(t)\) is monotonically decreasing instead. 

Furthermore, it can be seen from Eq.~\eqref{eq:chi_bound} that Eq.~\eqref{eq:magnus_criteria_tls} holds also 
for a bell-shaped driving. Indeed, assuming that \(f(t)\) is peaked at the origin and decays monotonically away from \(t=0\), by 
splitting the integration interval in two pieces, we find
\begin{eqnarray}
\int_{-\infty}^{t} E_{a}(\tau){\rm d}\tau&\leq& \frac{1}{2}|\chi(-\infty)-\chi(0)|+\frac{1}{2}|\chi(0)-\chi(t)|
\nonumber 
\\
&<&\pi.
\end{eqnarray}

Thus, the Magnus expansion can be safely applied to the non-adiabatic transition problem for a large class of drivings. For example, the LZSM driving \(f(t)=vt\) is monotonic over the whole time domain, while the Rosen-Zener\cite{RosenZener1932, BLANES2009151} driving ${f}(t)\propto\rm{sech}(t)$ belongs to the bell-shaped type. When the driving function oscillates, the Magnus expansion can be still applied in a piecewise manner to successive time intervals. 

Once the  Magnus operator is determined, using Eq.~\eqref{eq:su2_exp},  its equivalence to Eq.~\eqref{eq:Ua-matrix} is established by the relations
\begin{align}
\alpha=&\cos{\theta_{a}} -i\frac{\sin\theta_{a}}{\theta_{a}}C_{a},
\label{eq:alpha}
\\
\beta=&-i\frac{\sin\theta_{a}}{\theta_{a}}A_{a}^*,
\label{eq:beta}
\end{align}
where \(\theta_{a}\) is given in Eq.~\eqref{eq:angle_axis_AC_1} while \(A_{a}(t)\) and \(C_{a}(t)\) are the coefficients obtained from the Magnus expansion in the adiabatic picture. The non-adiabatic transition probability then follows from
\begin{equation}
\label{eq:prob}
P(t)=|\beta(t)|^2=\left|\frac{\sin\theta_{a}}{\theta_{a}}\right|^2|A_{a}(t)|^2.
\end{equation}

To provide an intuition for this procedure, we briefly consider the leading order approximation for the general case with the initial time \(t_0=-\infty\). By substituting Eq.~\eqref{eq:v_adiabatic} into Eq.~\eqref{eq:A1}, we get
\begin{equation}\label{eq:A1_adiabatic}
    A_1(t)=i\int^t_{-\infty} \dot\chi(s) {\rm e}^{2i\varphi(s)}{\rm d}s.
\end{equation}
Meanwhile, in first-order Magnus approximation (FMA),
\begin{equation}
    A_{a}^{\text{FMA}}(t)= A_1(t), \ C_{a}^{\text{FMA}}(t)=0\Rightarrow \theta_{a}^{\text{FMA}}(t)=|A_1(t)|
\end{equation}
and thus Eqs.~\eqref{eq:alpha},\eqref{eq:beta} 
turn into
\begin{eqnarray}
\alpha^{\text{FMA}}(t)&=&\cos{|A_1(t)|}.\label{eq:alpha_1st}
\\
\beta^{\text{FMA}}(t)&=&-i\frac{\sin{|A_1(t)|}}{|A_1(t)|}A^*_1(t),\label{eq:beta_1st}
\end{eqnarray}
In the adiabatic limit, i.e., \(|A_1(t)|\to 0\), Eq.~\eqref{eq:beta_1st} can be further approximated by
\begin{equation}\label{eq:beta_FMA}
    \beta^{\text{FMA}}(t)\approx -i A_1^*(t)=-\int^t_{-\infty} \dot\chi(s) {\rm e}^{-2i\varphi(s)}{\rm d}s,
\end{equation}
which is also the starting point of the 
complex WKB approximation\cite{DavisPechukas1976}. 

Finally, we stress that Eq.~\eqref{eq:alpha_1st} 
contains no phase information about 
\(\alpha(t)\). In particular, the Stokes 
phase\cite{Kaya97,SHEVCHENKO20101}, defined as
\begin{equation}
    \varphi_S:=-\arg[\alpha(+\infty)],
\end{equation}
is crucial for the discussion of 
Landau–Zener–Stückelberg 
interference\cite{SHEVCHENKO20101,Forster2014},
however. We show below that \(\varphi_S\) 
can be extracted from the second 
(and higher) order Magnus expansion.

\subsection{Magnus expansion for the LZSM model}

As an illustrative example for its quick convergence, we now explicitly consider the Magnus expansion for the LZSM model, i.e., a linear drive of the form $f=vt$ with sweep velocity $v>0$. Its final non-adiabatic transition probability is given exactly analytically by the so-called Landau-Zener formula\cite{Landau1932,Zener1932}
\begin{equation}
P(+\infty)=\exp(-2\pi\gamma), 
\end{equation}
with the definition
\begin{equation}
\label{eq:gamma}
\gamma:=\frac{\Delta^2}{4v}.
\end{equation}
For small sweep velocities the above probability 
vanishes, which is a special case of a well-known general result \cite{Dykhne1962}. 
Furthermore, the Stokes phase reads
\begin{equation}
\varphi_S=\frac{\pi}{4} +\gamma(\ln\gamma -1)+\arg\left[\Gamma(1-i\gamma)\right].
\end{equation}
These exact results can, e.g., be found by converting the Schr\"odinger equation into a second-order ordinary differential equation and identifying the asymptotic behavior of its solution in terms of special functions.\cite{Zener1932, SHEVCHENKO20101}

We note that the LZSM model displays parity time-inversion (PT) symmetry, i.e.,
due to the fact that \(f(-t)=-f(t)\in\mathbb{R}\), it follows that
\begin{equation}
\label{eq:PT}
PH^*(-t)P= H(t),
\end{equation}
where $P=\sigma_z$. To explicitly maintain this symmetry for \(H_a(t)\) in the adiabatic picture, we choose  \(\varphi(0)=0\) for the dynamical phase defined in Eq.~\eqref{eq:dynamical_phase}. Under such a choice, we have 
\begin{equation}
    PH^*_a(-t)P=H_{{a}}(t).
\end{equation}
As shown in Appendix~\ref{sec:symmetry_app}, due to the PT symmetry, \(A_n^*(t_f=\infty)=-A_n(t_f=\infty)\) holds for each order of the ME, i.e., the $A$-coefficients
at infinity are purely imaginary. Furthermore, the dynamical phase in the LZSM case reads 
\begin{equation}
    \varphi(t)=\frac{1}{2}\int_{0}^t\sqrt{\Delta^2+v^2s^2}{\rm d}s=\gamma  g(vt/\Delta),
\end{equation}
where $g(x):=x\sqrt{1+x^2}+\ln|x+\sqrt{1+x^2}|$
is an auxiliary dimensionless function. In addition, from Eq.\ (\ref{eq:chi_adiabatic}), we find 
\begin{equation}
\dot\chi(t)=\frac{v}{\Delta}\frac{1}{1+(vt/\Delta)^2}
\end{equation}
in the LZSM case. For convenience, we employ a dimensionless time variable \(x\equiv vt/\Delta\), so that the complex function in Eq.~\eqref{eq:hamiltonian_magnus} can be written as (also in dimensionless form)
\begin{equation}
    v(x)=\frac{i{\rm e}^{i2\gamma g(x)}}{2(1+x^2)}.
\end{equation}

\begin{figure}[htbp]
    \begin{subfigure}[t]{0.4\textwidth}
       \label{fig:LZ_prob}
        \includegraphics[width=\textwidth]{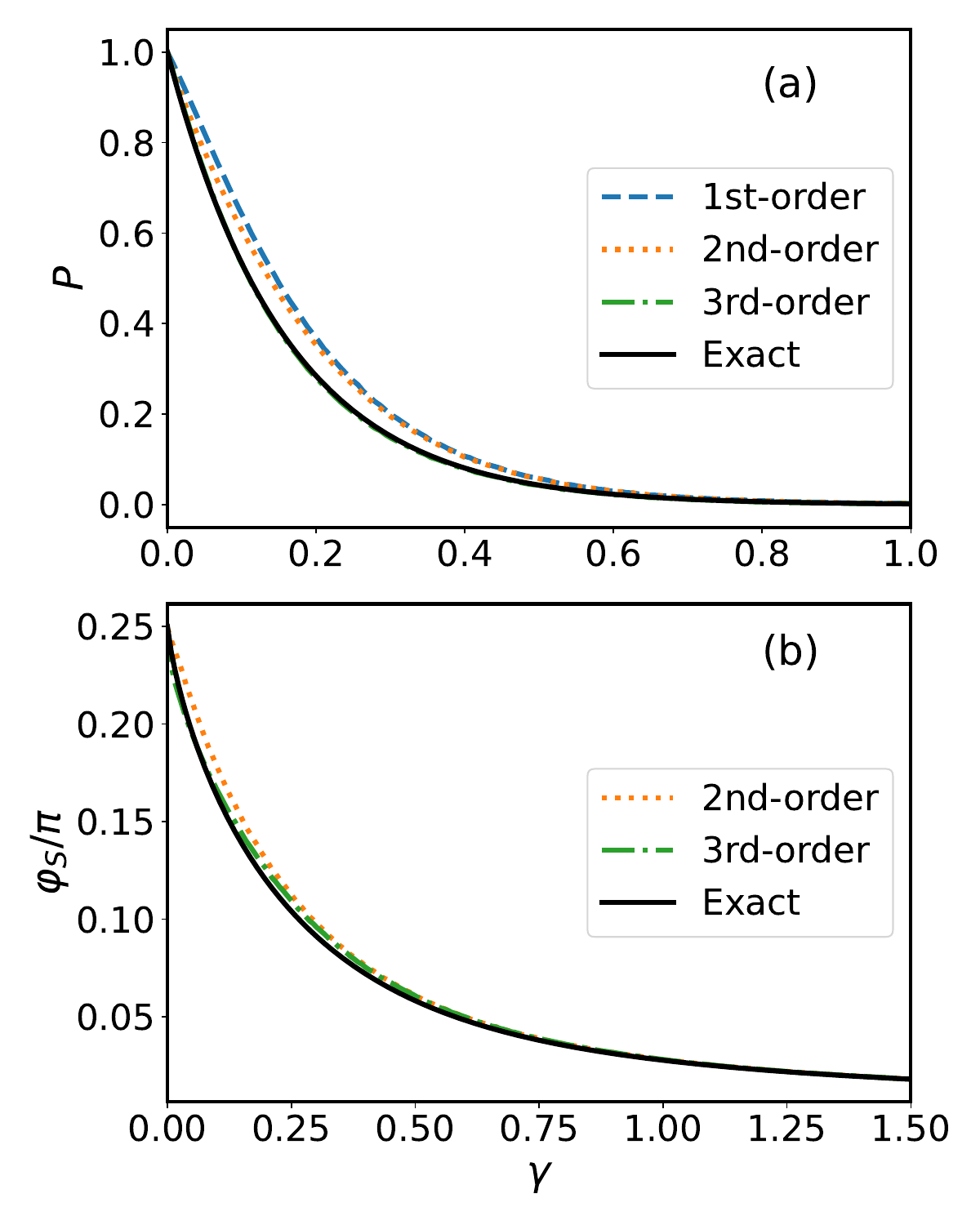}
    \end{subfigure}
    \caption{Magnus approximations of different order (1st order: dashed blue, 2nd order: dotted orange, 3rd order: dash-dotted green) and exact results (solid black) for the LZSM problem: (a) transition probabilities (exact and 3rd order results are almost indistinguishable), (b) Stokes phase (1st order not shown, as it gives a zero result) in units of $\pi$, both as a function of $\gamma$, defined in Eq.\ (\ref{eq:gamma}).} 
    \label{fig:Landau_Zener}
\end{figure}

In FMA we now finally have
\begin{equation}
 A_1(+\infty)= iJ,
\end{equation}
where \(J\) is determined by taking the limit \(t=+\infty\) in Eq.~\eqref{eq:A1_adiabatic} and it can be simplified as
\begin{equation}
\label{eq:Jofg}
J(\gamma)=\int_0^{+\infty}\frac{\cos[\gamma (\sinh(2s)+2s)]}{\cosh(s)}{\rm d}s.
\end{equation}
As shown in panel (a) of Fig.~\ref{fig:Landau_Zener},
the FMA yields qualitatively good results for the transition probability calculated from Eq.\ (\ref{eq:prob}) together with Eq.\ (\ref{eq:beta_1st}). In order to uncover the Stokes phase, we have to consider at least the second order Magnus approximation (SMA), however. It then follows from the resolution for the angle of
the transcendental equation
\begin{equation}
\tan(\varphi_{S}^{\text{SMA}})=\frac{\tan\sqrt{|A_1|^2+|C_2|^2}}{\sqrt{|A_1|^2+|C_2|^2}}C_2,
\end{equation}
where
\begin{equation}\label{eq:C2_LZ}
    C_2(\gamma)=\iint_{x_2<x_1}\frac{\sin(2\gamma [g(x_1)-g(x_2)])}{(1+x^2_1)(1+x_2^2)}{\rm d}x_1{\rm d}x_2.
\end{equation}
according to Eq.~\eqref{eq:C2}. 

For a graphical display of the results, 
in a final step, the integrals in Eqs.\ 
(\ref{eq:Jofg}) and (\ref{eq:C2_LZ}) 
and analogous expressions for higher order have to be performed numerically.
The corresponding "semi-analytical" 
results up to third order for the 
transition probability as well as the 
Stokes phase are depicted in 
Fig.~\ref{fig:Landau_Zener}. 
Remarkably, the 
third order approximation already yields 
almost exact values for the transition 
probability over the whole range of $\gamma$
that is displayed and a little less accurate results for the Stokes 
phase. 
Furthermore, when \(\gamma\) is either large or small, also the lower ME orders are quite accurate, i.e., we do not encounter the celebrated \(\pi/3\)-problem that appears if the integration in Eq.\ \eqref{eq:beta_FMA} is done analytically in the complex plane in first order
\cite{DavisPechukas1976}. The former case can be understood by noticing that the integrand in Eq.\ (\ref{eq:C2}) oscillates faster as \(\gamma\) increases, and thus higher order Magnus terms vanish. While when \(\gamma\) is small, \(H_a(t)\) varies slowly with time and thus commutators of it at different times are small. In the sudden limit \(\gamma\to 0^+\), \(A_1(\gamma)\) and \(C_2(\gamma)\) yield the exact value \(\varphi_S=\pi/4\). 

\section{Floquet Quasienergies 
and the Rabi model}\label{sec:Floquet}

The Floquet quasienergy \(\epsilon\)  determines the long-time evolution of a periodically driven system. Considering the case \(f(t+T)=f(t)\) in Eq.~\eqref{eq:phys_hamiltonian} with period \(T\), in his seminal work, Shirley \cite{Shirley1965} applied the Floquet formalism and the Hellmann-Feynman theorem (see Eq. 24 in~\cite{Shirley1965}) to prove the following result for the time-averaged transition probability from level one to level two:
\begin{equation}\label{eq:shirley_fh}
    \bar{P}_{1\rightarrow2}=\frac{1}{2}\left[1-4\left(\frac{\partial \epsilon}{\partial\Delta}\right)^2\right]. 
\end{equation}
This result allows us to directly gain a physical understanding from the quasienergy spectrum as a function of \(\Delta\). For example, the resonance condition is located at avoided crossings since \(
\frac{\partial \epsilon}{\partial\Delta} =0.
\) At exact crossings, however, transition probabilities may become zero~\cite{CDT-PRL-1991}. Furthermore, in their vicinity, the dissipative behavior is sensitive to small parameter variations \cite{Engelhardt2019,Kohler2024}.

 As reviewed in \cite{FGbook}, the Floquet quasienergy 
\(\epsilon\)
can be extracted from the phase of 
the time-evolution operator over one period, \(U(T)\), independent of the choice of initial 
time \(t_0\). For example, in the two-level case, we have 
\begin{equation}
\label{eq:qen}
\epsilon=\pm \frac{\theta(T)}{T}=\pm \frac{\theta(T)}{2\pi}\omega \mod{\omega},
\end{equation}
where \(\omega=2\pi/T\) denotes the frequency. In the remainder of this presentation, we will omit writing \(\pm\) and mod for \(\epsilon\) since there is only one independent quasienergy due to the traceless nature of the Hamiltonian under study. 

Alternatively, the quasienergy corresponds to an eigenvalue of the effective 
Hamiltonian, which is defined modulo \(2\pi/T\),
\begin{equation}
H_{\text{eff}}\equiv \frac{1}{T} \Omega(T).
\end{equation}
 First applications of the Magnus approximation
to periodically driven quantum dynamics have
been reported in the field of nuclear magnetic resonance \cite{Haeberlen1968,Evans1968,Feldman1984}. Despite the fact that the connection between the quasienergy and the Magnus operator appears quite natural, and the so-called Floquet-Magnus expansion\cite{BLANES2009151,Zeuch2020}  has been discussed and employed to compute \(H_{\text{eff}}\), only few authors have used the above connection to compute the quasienergy spectrum \cite{Lopez2013graphene}. 

 In the following, we first propose a scheme for calculating the quasienergies using suitable picture transformations together with the Magnus expansion. This scheme is, in principle, valid for a general driving shape and any coupling strength. However, we note that to properly distinguish exact and avoided crossings, further considerations about symmetry have to be taken into account. Our methodology is then illustrated by its application to the semiclassical Rabi model.

\subsection{Quasienergy computation via Magnus expansion}

Our strategy to compute the quasienergy is to perform the Magnus expansion in the picture appropriate for the  studied parameter regime. To make progress, we first isolate the effect of the driving strength by rewriting the 
function \(f(t)\) in Eq.~\eqref{eq:phys_hamiltonian} as 
\[
f(t)\equiv g\tilde{f}(t), \quad g>0, \quad  \tilde{f}(t+T)=\tilde{f}(t), \quad \left|\tilde{f}(t)\right |\leq 1,
\]
where \(g\) denotes the driving strength and thus the shape function \(\tilde{f}(t)\) is dimensionless. With \(\tilde{f}(t)\) given, there are only two independent dimensionless parameters, which can be chosen as \(g/\omega\) and \(\Delta/\omega\).  For notational convenience, in the following we choose \(\omega=1\) and thus \(T=2\pi\). To recover the dimension, one uses the correspondence
\begin{equation}
    \epsilon = \epsilon\left(\frac{\Delta}{\omega}, \frac{g}{\omega}\right)\omega .
\end{equation}

\begin{figure}[hbtp]
\centering
\includegraphics[width=0.6\linewidth]
{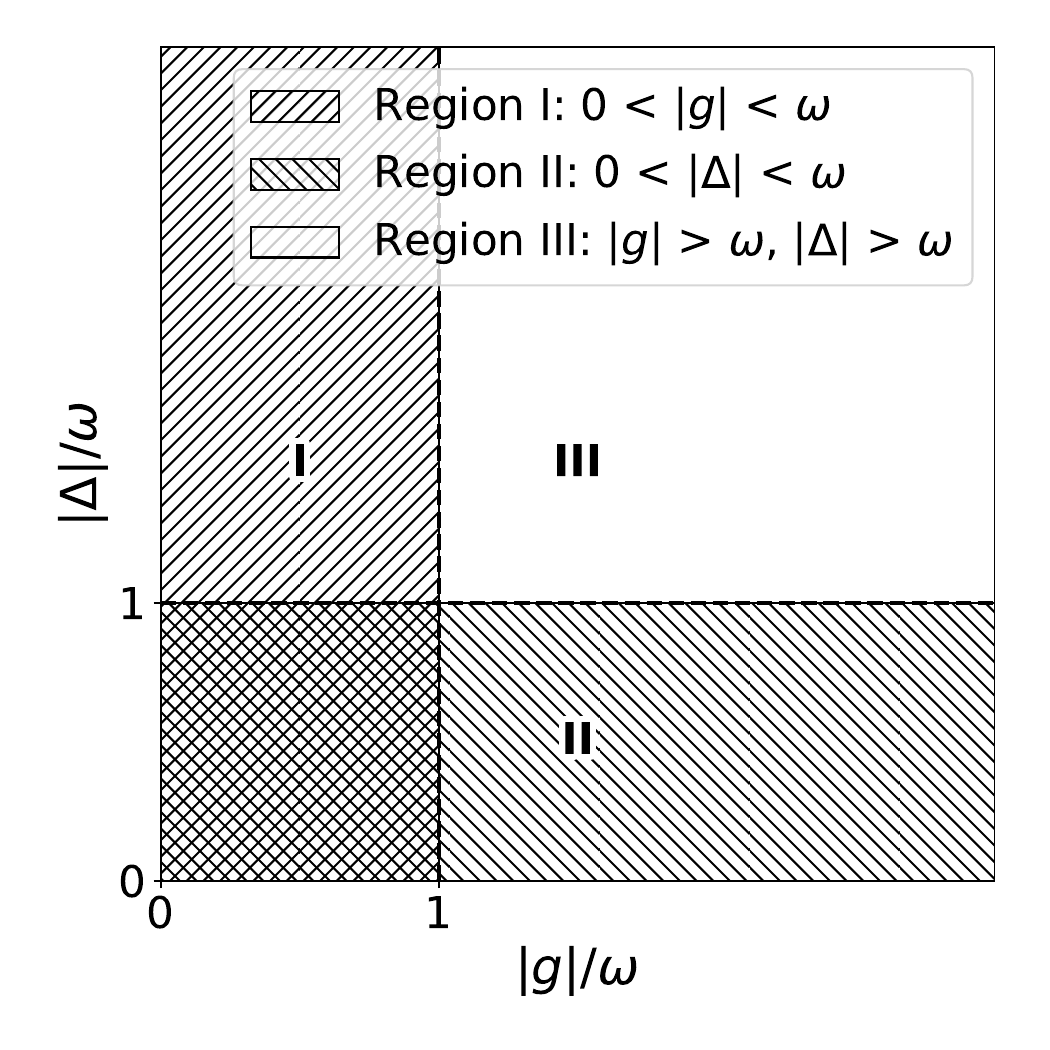}
\caption{Parameter regions for which the Floquet quasienergies of the semiclassical Rabi model are 
investigated.}
\label{fig:para}
\end{figure}

\begin{table}
\begin{ruledtabular}
\begin{tabular}{cccc}
   &I: $g<\omega$ &II: $\Delta < \omega$&III: $g>\omega, \Delta>\omega$
\\
\hline
\\
$v(t)$ & $\frac{g}{2}\tilde{f}(t){\rm e}^{-i\Delta t}$ & $\frac{\Delta}{2}{\rm e}^{i{F}(t)}$ & $\frac{1}{2}\dot\chi(t) {\rm e}^{-i2\varphi(t)}$\\
\end{tabular}
\end{ruledtabular}
\caption{\label{tab:table1}Parameter regimes and the 
corresponding functions \(v(t)\) that enter the interaction picture version of Eq.~\eqref{eq:hamiltonian_magnus}, which is employed for computing the Floquet quasienergy via the Magnus expansion.}
\end{table}
 Next, we divide the entire \((g, \Delta)\) plane into three parameter regimes as indicated in Fig.\ \ref{fig:para} and employ suitable picture transforms such that the ME can be safely applied. Our scheme is summarized in Table~\ref{tab:table1}, where \(v(t)\) is the crucial ingredient of the Hamiltonian. For each parameter region, \(v(t)\) is chosen such that the instantaneous energy \(E(t)\equiv|v(t)|\), corresponding to the operator norm, is small so that the ME exhibits controllable convergence properties. 

\subsubsection*{Region I: \(|g| <\omega\)}

Let us first consider the case 
\(| g| <1=\omega\), where we regard the 
off-diagonal part of 
Eq.~\eqref{eq:phys_hamiltonian} as the 
perturbation and transform into the 
corresponding interaction picture. More 
specifically, we let \(H(t)=H_0+V(t)\) with
\begin{equation}
H_0=\frac{\Delta}{2}\sigma_z, \ V(t)=\frac{g\tilde{f}(t)}{2}{\sigma_x}.
\end{equation}
For the transformation into the interaction
picture, we  use the unitary transformation
\begin{equation}\label{eq:U0_g}
     U_0(t)=\exp\left(-i\frac{\Delta t}{2}\sigma_z\right)
\end{equation}
Furthermore, the evolution operator 
\(U_I(t)=U_0^{-1}(t)U(t)\) is generated by the Hamiltonian 
\begin{equation}\label{eq:HI_g}
    H_I(t)=U_0(t)^{-1} V(t)U_0(t)=v_g(t)\sigma^++v^*_g(t)\sigma^-,
\end{equation}
with 
\begin{equation}
v_g(t)=\frac{g}{2}\tilde{f}(t)
{\rm e}^{i\Delta t},
\end{equation}
which is the entry in the first column of Table \ref{tab:table1}.
Since \(|g|<1\), the convergence condition Eq.~\eqref{eq:magnus_criteria_tls} is satisfied for 
the whole interval \(t\in [0, 2\pi]\):
\begin{equation}
\int_0^{t} |v_g(s)|{\rm d}s
\leq
\int_0^{t} \frac{|g|}{2}{\rm d}s
=\frac{|g|t}{2} <\pi.
\end{equation}
This inequality holds without any condition for the magnitude of \(\Delta\).

In order to find the angle-axis parameterization in Eq.~\eqref{eq:U_angel_axis}, one needs to solve the BCH problem\cite{GilmoreBCH1974}, i.e., to use 
\begin{equation}
\exp(-i\theta \vec n \cdot \vec \sigma)=\exp(-i\theta_0 \vec n_0 \cdot \vec \sigma)\exp(-i\theta_I \vec n_I \cdot \vec \sigma)
\end{equation}
and to determine \(\theta\) and \(\vec{n}\) from
the pairs \((\theta_0,\vec{n}_0)\) and \((\theta_I,\vec{n}_I)\).
Using the properties of the Pauli matrices, we get
\begin{align}
    \cos{\theta}&=\cos\theta_0\cos\theta_I-\sin\theta_0\sin\theta_I  \vec n_0 \cdot \vec n_I,\label{eq:BCH_a}
    \\
  \sin{\theta}\ \vec n  &=\sin\theta_0\cos\theta_I  \vec n_0+\cos{\theta_0}\sin\theta_I \vec n_I\notag\\&+\sin\theta_0\sin\theta_I\vec n_0\times \vec n_I.
  \label{eq:BCH_b}
\end{align}
\(\theta_0\) and \(\vec n_0\) will be read off from the unitary transformation matrix \(U_0\). For example, from Eq.~\eqref{eq:U0_g} one finds
\begin{equation}\label{eq:theta_n_0_g}
    \theta_0(t)=\frac{\Delta t}{2}, \quad \vec{n}_0=\hat{e}_z.
\end{equation}
\(\theta_I\) and \(\vec n_I\), on the other hand,  will be determined from the ME, i.e., from the insertion of Eq.\ (\ref{eq:AC}) into Eqs. (\ref{eq:angle_axis_AC_1},\ref{eq:angle_axis_AC_2}). 

 Once we obtain \(A_I(2\pi)\) and \(C_I(2\pi)\) as well as \(\theta_I(2\pi)\) from the ME for \(U_I(2\pi)\), using Eq.~\eqref{eq:BCH_a} together with 
Eq.~\eqref{eq:qen}, the quasienergy follows  to be
\begin{align}
    \epsilon_{2\pi}
& = \frac{1}{2\pi} \cos^{-1}{\left[\cos\theta_0 \cos\theta_I - C_I\sin\theta_0 \frac{\sin\theta_I}{\theta_I} \right]}.\label{eq:eps_2pi} 
\end{align}
The quasienergy is thus given in terms of easily accessible trigonometric and inverse trigonometric functions and the sinc function. Here, we explicitly denote the fact that the quasienergy is extracted from the time-evolution operator over the full period by the index $2\pi$. 

\subsubsection*{Region II: \(|\Delta|<\omega\)}

Next we consider the regime \(|\Delta|<1\). To 
reuse the above derivation, we exchange 
\(\sigma_x\) and \(\sigma_z\) in 
Eq.~\eqref{eq:phys_hamiltonian} by the 
2x2 Hadamard matrix \(H_2\equiv\frac{1}{\sqrt{2}}(\sigma_x+\sigma_z)\), via
\begin{equation}\label{eq:switch_xz}
    H_2\sigma_xH_2=\sigma_z,\ H_2\sigma_zH_2=\sigma_x.
\end{equation}
This is a time-independent similarity transform and does not affect any physical observables, as well as the quasienergy. Compared to the previous regime \(|g|<1\), we switch also \(H_0\) and \(V\), namely we set
\begin{equation}
\ H_0(t)=\frac{g\tilde{f}(t)}{2}\sigma_z, \ V=\frac{\Delta}{2}\sigma_x,
\end{equation}
and stress that the unperturbed Hamiltonian is now time-dependent, although it does not contain non-commuting operators. Thus a transformation into the interaction picture using
\begin{align}
    U_0(t)&=\exp\left(-i\frac{g \tilde{F}(t)}{2}\sigma_z\right)
  \label{eq:U0_d}
\end{align}
with
\begin{equation} 
\label{eq:defF}
\tilde{F}(t)\equiv \int_0^t \tilde{f}(s)ds
\end{equation}
yields
\begin{equation}\label{eq:HI_d}
H_I(t)=v_\Delta(t)\sigma^++v^*_\Delta(t)\sigma^-, 
\end{equation}
where
\begin{equation}
\label{eq:vDelta}
v_\Delta(t) =\frac{\Delta}{2}{\rm e}^{ig\tilde{F}(t)}.
\end{equation}
Again, \(|\Delta|<1\) ensures that the ME for \(U_I(t)\) converges within a period. 

The quasienergy can be then determined by again using Eq.~\eqref{eq:eps_2pi}, but now with
\begin{equation}\label{eq:theta_n_0_d}
    \theta_0(t)=\frac{g}{2}\tilde{F}(t),
\end{equation}
instead of the use of Eq.~\eqref{eq:theta_n_0_g}.

\subsubsection*{Region III: \(\omega < |g|\) and \(\omega <|\Delta|\)}

The last parameter regime left to consider is given by \(\omega < |g|\) and \(\omega <|\Delta|\), which intuitively means the driving is relatively slow-varying. We expect the adiabatic picture explained in Sec.~\ref{sec:non-adiabatic} to be a suitable choice. We repeat Eq.~\eqref{eq:U_adiabatic_pic}, but now for the time-evolution over one period,
\begin{equation}
U(T,t_0=0)=\Psi(T)\Phi(T)U_a(T)\Phi^\dagger(0)\Psi^\dagger(0),
\end{equation}
where the definition of \(U_0\) in Eq.~\eqref{eq:psi_a} has been inserted. Since the Hamiltonian is periodic, we have \(\Psi(T)=\Psi(0)\). Furthermore, \(\Phi(T)\) contributes the dynamical phase, whereas under the adiabatic approximation \(U_a(T)\) is only responsible for the geometric, so-called Berry phase\cite{BerryPhase1984}. Thus, in the adiabatic limit, we have
\begin{equation}
\label{eq:Berry}
\epsilon =\frac{1}{T}[\varphi(T)+\varphi_{\text{Berry}}(T)]
\end{equation}
for the quasienergy. The geometric contribution for single-axis driving 
considered here is zero. 
For a more detailed discussion of the dynamical/geometrical decomposition of the Floquet quasienergy in the more general case, we refer the reader to\cite{SchindlerPRX2025}. In the following, we apply the ME to the Rabi model and at the end of the next subsection, we show 
how to go beyond the adiabatic approximation.
\subsection{Semiclassical Rabi Model}\label{sec:Rabi}

In the present section we demonstrate the 
application of the scheme outlined above by studying the semiclassical Rabi model, where the shape function \(\tilde{f}(t)\) can be chosen either as \(\cos{\omega t}\) or \(\sin{\omega t}\). The concrete choice does not affect the {\it exact} analytic/numeric value of the quasienergy and we will be using
\begin{equation}
    \tilde{f}(t)=\cos\omega t
\end{equation}
in the following.
Yet we would like to point out that employing \( \tilde{f}(t)=\sin\omega t\), instead, may lead to qualitative differences for the ME, since sine and cosine functions possess different symmetries in the interval \([0, \pi]\), see also Appendix~\ref{sec:symmetry_app}. 

For either choice, the Hamiltonian Eq.~\eqref{eq:phys_hamiltonian} obeys the so-called generalized parity (GP) symmetry
\begin{equation}
    PH(t+\pi)P=H(t),
\end{equation}
where \(P=\sigma_z\). The relevance of this 
symmetry in the context of periodically 
driven tunneling was highlighted in 
\cite{zpb91}, as well as in \cite{Perez1991}. In \cite{zpb91}, based on a classic argument by von Neumann and Wigner \cite{vonNeumann1929}, it was shown that GP symmetry 
allows for exact crossings of quasienergies along one-dimensional 
manifolds in a two-dimensional parameter space made up of frequency 
and driving strength \cite{dissown}. In the 
two-level system studied herein similar crossings may occur \cite{epl92, PRB_crossing_2003}.
The two parameters to be varied herein are driving strength $g$ and $\Delta$, i.e., the unperturbed level splitting, however. By using the GP symmetry one can prove 
that \(\epsilon\) is an odd function of \(\Delta\) and an even function of \(g\), namely,
\begin{equation}\label{eq:Rabi_eps_symmetry}
     \epsilon(\Delta, g)= \epsilon(\Delta, -g)=- \epsilon(-\Delta, g).
\end{equation}
Thus we may focus on the case where all parameters are positive valued. However, as will be seen below, the existence of the GP symmetry also leads to some subtleties that require special attention. 

\subsubsection*{Exact analytic results} 
\begin{figure}[htbp]
\includegraphics[width=0.48\textwidth]{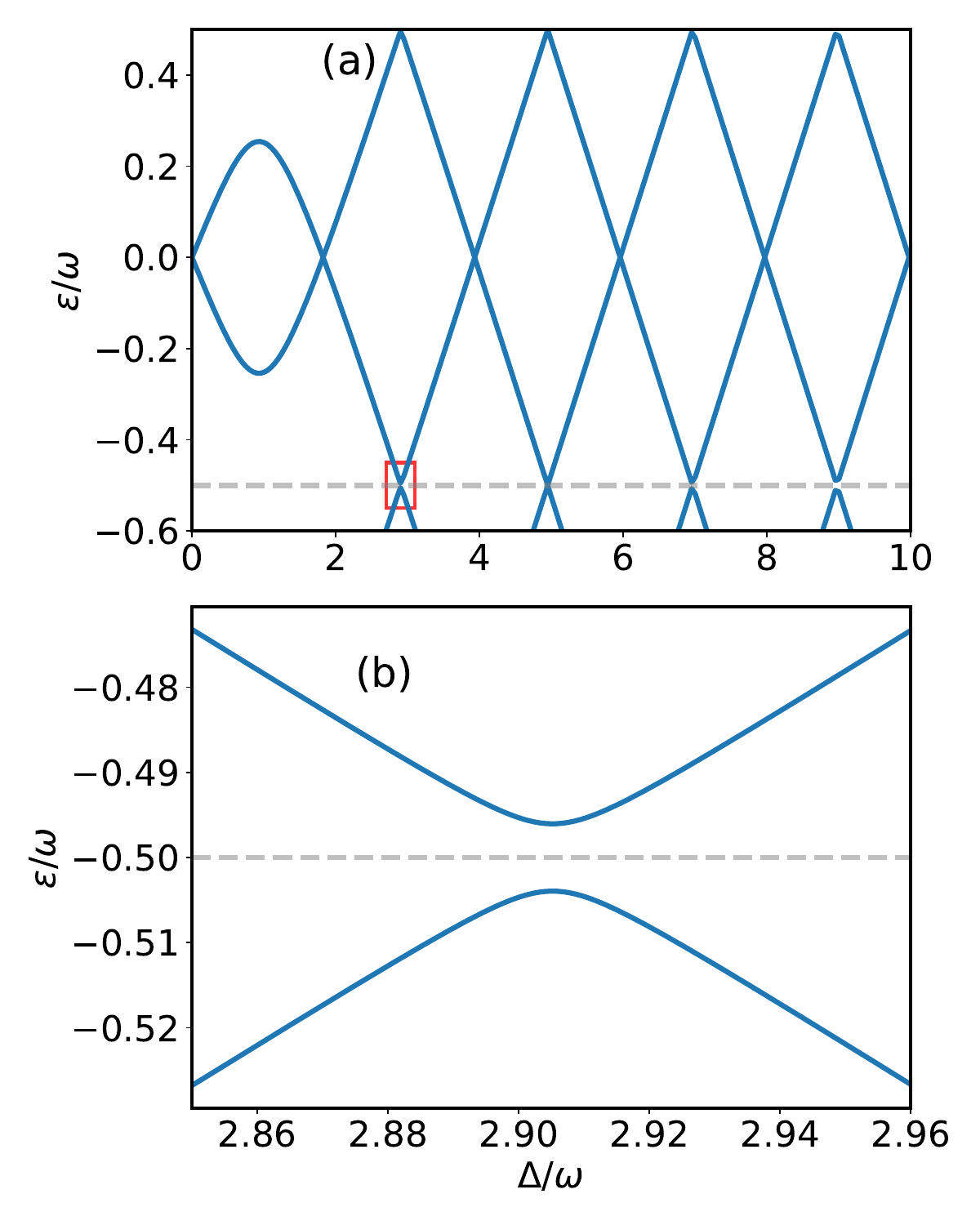}
    \caption{Exact results for the quasienergy of the Rabi model at \(g/\omega=1\) as a function of $\Delta$. Panel (b) shows a magnified view for the boxed region in panel (a), displaying an avoided crossing.}
    \label{fig:eps_exact}
\end{figure}

As a reference to compare our ME results to,
we reproduce the exact solution for the quasienergy, given by Schmidt and collaborators 
\cite{Schmidt2018, Schmidt2019}. It was found that
\begin{equation}\label{eq:rabi_quasi_energy}
    \epsilon(\Delta, g)=\frac{1}{\pi}\sin^{-1}{\left[\sqrt{2}\Delta\text{Re}
    \left({\rm e}^{ig}\eta_+(g, \Delta)
    \eta_-(g, \Delta)\right)\right]}
\end{equation}
holds for the quasienergy, where
\begin{align*}
   & \eta_\pm(\Delta, g):=\text{HeunC}(\pm\mu_0, \mu_1, b_0, b_1, a;z=\frac{1}{2}),\\
    &\mu_0=\mu_1=\frac{1}{2}, \quad a = 2ig, \\
    &b_0 = -\frac{1}{8}(4ig+2\Delta^2+1),  \quad b_1 = ig.  
\end{align*}
Here "HeunC" denotes Heun's confluent function\cite{NIST:DLMF}, which is implemented in standard mathematical
software packages and/or can be programmed via Taylor expansion. In Fig.~\ref{fig:eps_exact} the exact analytical quasienergy \(\epsilon(\Delta, g=1)\) is depicted as a function of \(\Delta\). As shown by the magnifying view in panel (b) of the figure, near the boundary of the first Brillouin zone (extending from -0.5 to 0.5) crossings are avoided, as to be expected from the argument given in Appendix~\ref{sec:crossing_app}. 

We note in passing that the exact quasienergies can
also be calculated numerically by diagonalizing the so-called Floquet matrix \cite{FGbook}. Especially for large values of $g$, the number of 2x2 sub-matrices to be taken into account to achieve convergence of the final result increases considerably, however.

\subsubsection*{Region I}

Applying the ME, firstly, we consider the weak driving case {I}, i.e., \(|g|<\omega\). We recall that the quasienergy follows from Eq.~\eqref{eq:eps_2pi} once  \(C_I\) and  \(\theta_I\) are obtained for  \(U_I(T)\). 

\begin{figure}[htbp]
        \centering
\includegraphics[width=0.45\textwidth]{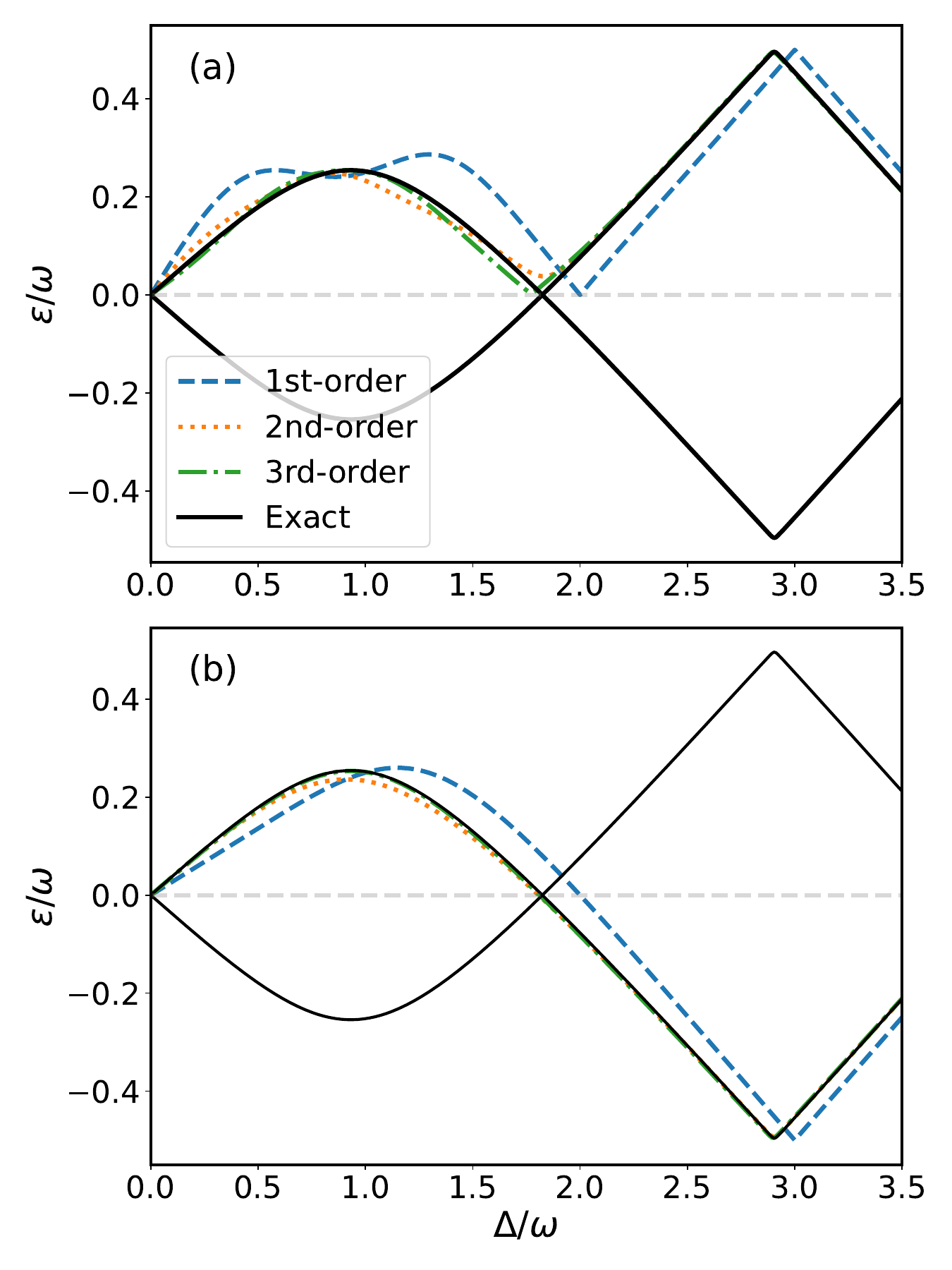}
    \caption{Quasienergies for \(g/\omega=1\): (a) exact result for $\epsilon$ and $-\epsilon$ (solid black lines) and Magnus approximations of different order based on the picture appropriate for region I
    (1st order: dashed blue, 2nd order: dotted orange, 3rd order: dash-dotted green)
    for $\epsilon$, directly computed from \(U(2\pi)\); (b) same as (a) but with Magnus approximations with generalized parity symmetry explicitly maintained by using \(U(\pi)\) for the determination of $\epsilon$ (see text). }
    \label{fig:eps_g}
\end{figure}

 Fig.~\ref{fig:eps_g} (a) shows the results for the first three orders of the Magnus approximation for \(g=1\) at the border of the first domain and up to $\Delta/\omega\approx 3$, i.e., up to the avoided crossing highlighted in Fig. \ref{fig:eps_exact} (b). For smaller \(|\Delta|\) the ME converges faster and the results are more accurate. As the order increases, the accuracy does improve. 
In particular, the second order 
is needed to observe the shift of the position of the avoided crossing, which corresponds to the Bloch-Siegert shift\cite{BS-shift1940},
originally introduced in nuclear magnetic resonance and discussed in the present
context by Shirley \cite{Shirley1965}. 

There is an apparent flaw in the results
presented in panel (a) of Fig.\ \ref{fig:eps_g}, however. This is the fact that the {\it exact} crossings at \(\Delta/\omega\approx 1.82\) (see Fig.\ \ref{fig:eps_exact}(a)) are artificially {\it avoided}
by the Magnus results. We note in passing that a similar issue occurs for another related model\cite{Lopez2013graphene}.
This flaw arises because the GP symmetry involves a time-translation operation. Unlike PT symmetry, it is artificially violated under finite order Magnus approximation due to truncation errors. 

To fix the issue, we have to explicitly maintain the GP symmetry. It has been noticed in~\cite{Schmidt2019} that, in the presence of this symmetry, the evolution operator over the whole period can be expressed by the one over half a period according to
\begin{equation}
U(2\pi)=\left[PU(\pi)\right]^2.
\end{equation}
As discussed in Appendix~\ref{sec:crossing_app}, this relation makes the exact crossings at \(\epsilon=0\) much easier to find. Motivated by this,  we will extract the quasienergy from the Magnus approximation to the time-evolution operator $U(\pi)=U_0(\pi)U_I(\pi)$ over only half a period. In Appendix~\ref{sec:crossing_app} we prove that,
\begin{equation}\label{eq:eps_from_GPtr}
   \text{tr}[PU(\pi)]=\sin{(\epsilon \pi)}=\vec{n}_P\cdot(\sin{\theta}\ \vec{n})
\end{equation}
holds, where \(\vec{n}_P=\hat{e}_z\) is determined by \(P=\vec{n}_P\cdot \vec\sigma\). The remaining task is thus to determine \(\theta(\pi)\) and \(\vec{n}(\pi)\).

To this end, firstly, from Eq.~\eqref{eq:U0_g} we read off
\begin{equation}\label{eq:theta_0_g_pi}
    \theta_0(\pi)=\frac{\Delta \pi}{2}, \ \vec{n}_0=\hat{e}_z.
\end{equation}
Secondly, we have to determine \(\theta_I(\pi)\) and \( \vec{n}_I(\pi)\) 
for \(U_I(\pi)\) via the ME according to Eqs.~(\ref{eq:angle_axis_AC_1},\ref{eq:angle_axis_AC_2}). Then applying \(\hat{e}_z \cdot\) to both sides of Eq.~\eqref{eq:BCH_b} we have
\begin{align}\label{eq:np_sin_n_g}
    \hat{e}_z\cdot(\sin{\theta}\ \vec{n})&= \cos\theta_I\sin\theta_0 +\cos{\theta_0}\sin\theta_I \hat{e}_z\cdot\vec n_I.
\end{align}
 Finally, comparing Eq.~\eqref{eq:np_sin_n_g} with Eq.~\eqref{eq:eps_from_GPtr}, we get
\begin{equation}
\label{eq:half}
\epsilon_\pi =\frac{1}{\pi}\sin^{-1}{\left[
\sin \theta_0\cos\theta_I
+
C_I\cos\theta_0\frac{\sin\theta_I}{\theta_I}\right]},
\end{equation} 
where $\theta_0(\pi)$ is given in Eq.~\eqref{eq:theta_0_g_pi} while  \(C_I\) and \(\theta_I\) are now also taken at time $t=T/2=\pi$.
The complexity of the result is similar to the one 
of Eq.\ (\ref{eq:eps_2pi}) and 
the results gained from the improved formula are depicted in panel (b) of Fig.~\ref{fig:eps_g}. As expected, now
the exact crossings are properly reproduced
and in addition also the overall accuracy has improved due to the extra work. The second order results, e.g., are so well-converged that they barely can be distinguished from the exact ones in 
panel (b) over the whole range of $\Delta$. Going to third order the results are graphically almost indistinguishable from the exact ones.

\subsubsection*{Region II}
Secondly, we compute the quasienergy for the parameter regime of region {II}, i.e., \(|\Delta|<\omega\) with the corresponding identification of $v(t)=v_\Delta(t)$ given in the second column of Table \ref{tab:table1}. Specifically, with  \(\tilde{f}(t)=\cos{(t)}\) yielding \(\tilde{F}(t)=\sin{(t)}\), we get
\begin{equation}
    v_{\Delta}(t)=\frac{\Delta}{2}
    {\rm e}^{i\sin t}.
\end{equation}

For the discussion of the FMA in the present case we start with the time-evolution operator over the full period. We note that here \(\theta_0(2\pi)=0\) according to Eq.~\eqref{eq:theta_n_0_d} and hence \(H_I(t)\) remains periodic, which is a direct result by the GP symmetry. Thus from Eq.\ (\ref{eq:qen}), we can directly read off
\begin{equation}
\epsilon_{2\pi}=\frac{1}{2\pi}\theta_I(2\pi).
\end{equation}
Using Eqs.\ (\ref{eq:angle_axis_AC_1}) and (\ref{eq:A1}), 
the first order Magnus approximation thus generally yields
\begin{align}
\label{eq:eps_1st_d_2pi}
\epsilon_{2\pi}^{\text{FMA}}&=
\frac{|A^{\text{FMA}}_{I}(2\pi)|}{2\pi}=\frac{\Delta}{4\pi}\left|\int_0^{2\pi}\exp\left[ig\tilde{F}(\tau)\right]{\rm d}\tau\right|,
\end{align}
which under cosine-driving becomes
\begin{align}
\epsilon_{2\pi}^{\text{FMA}}
=\frac{\Delta}{2}
\frac{1}{\pi}
\left|\int_0^{\pi}\cos(g\sin\tau){\rm d}\tau\right|={\frac{\Delta}{2}J_0(g)}.
\end{align}
This is the zeroth-order Bessel function\cite{Abramowitz1964}
expression for the quasienergy of the semiclassical Rabi model, a result which has been found 
using perturbation theory more than sixty years ago by Shirley \cite{Shirley1965}
and can also be extracted from the Heun function solution in the limit $\Delta\to 0$\cite{Schmidt2019}. A physical interpretation of Eq.~\eqref{eq:eps_1st_d_2pi} has been given in terms of an effective tunneling rate that vanishes at the zeros of $J_0$ \cite{CDT-PRL-1991,epl92,Grifoni1998,Sierra-PRA-2015}. 

\begin{figure}[htbp]
    \centering
\includegraphics[width=0.9\linewidth]{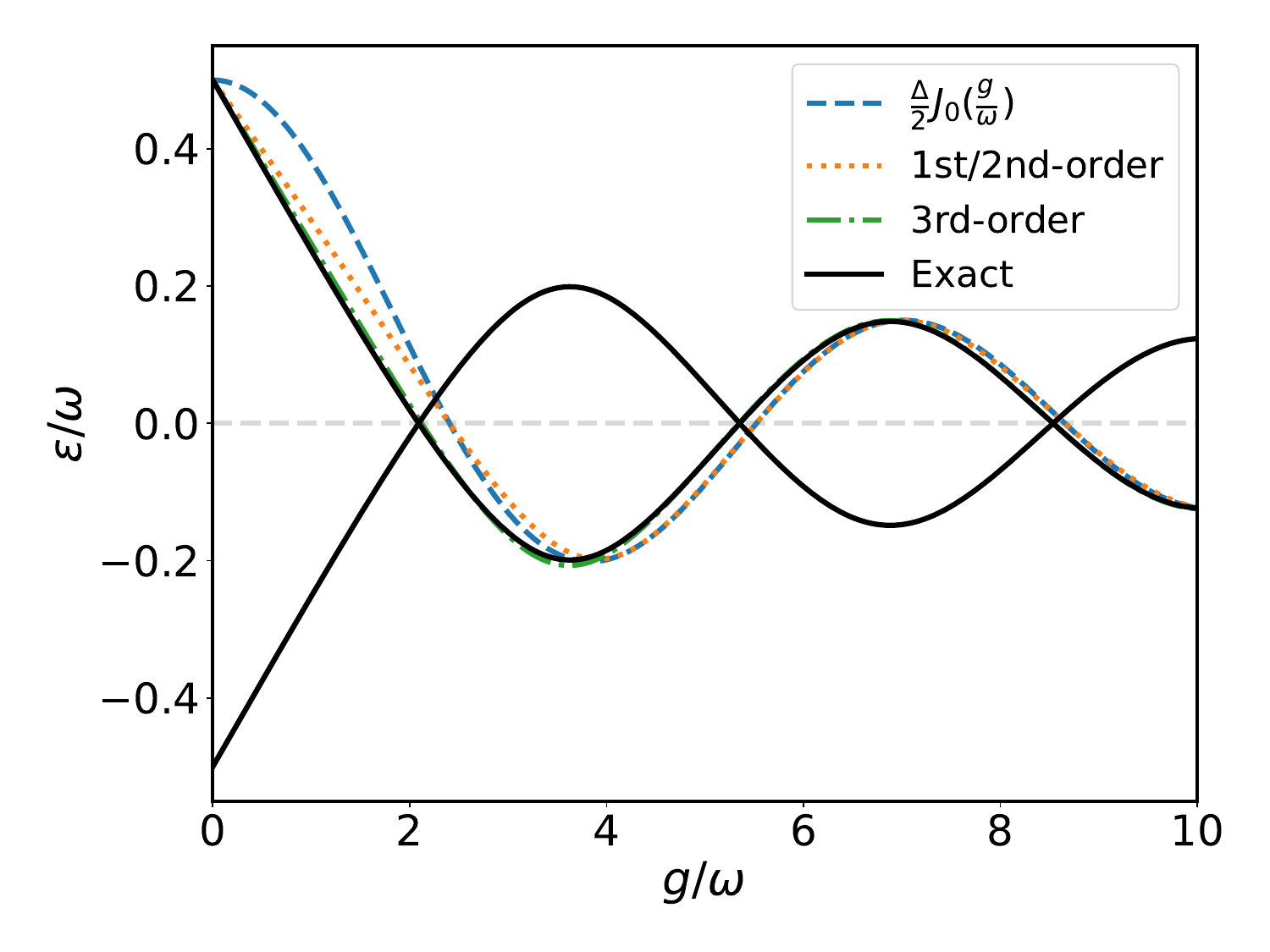}
    \caption{Quasienergy of the semiclassical Rabi model at \(\Delta/\omega=1\) as a function of $g$: exact result for $\epsilon$ and $-\epsilon$ (solid black lines) and Magnus approximations of different order based on the picture appropriate for region II: 1st order extracted from time evolution over $2\pi$: dashed blue, 
    1st (equal to 2nd) order with the generalized parity symmetry explicitly maintained: dotted orange, 3rd order with the generalized parity symmetry explicitly maintained: dash-dotted green.}
    \label{fig:eps_d}
\end{figure}

With the ME it is straightforward to find next order correction terms, yet we refrain from going to higher order for \(U_I(2\pi)\) here and turn to the approach with the GP symmetry explicitly maintained. To make progress, we note that we have exchanged \(\sigma_x\) and \(\sigma_z\) by Eq.~\eqref{eq:switch_xz} for region {II} and thus the representation of the parity operator changes accordingly, i.e.,
\begin{equation}
    P=\sigma_x, \quad \vec n_P=\hat{e}_x.
\end{equation}
In the present case we have 
\(\theta_0(\pi)=0\) and thus Eq.~\eqref{eq:BCH_b} simplifies considerably and the corresponding quasienergy reads
\begin{equation}
\label{eq:epsII}
\epsilon_\pi = \frac{1}{\pi}\sin^{-1}{\left[{\rm Re}(A_I) \frac{\sin\theta_I}{\theta_I}\right]}.
\end{equation}
Herein, the  \(A_I\) and \(\theta_I\) are again taken at $t=\pi$
as indicated by the quasienergy index. To compare Eq.~\eqref{eq:eps_1st_d_2pi} with Eq.~\eqref{eq:epsII}, for the latter the first order ME contribution reads
\begin{align}
   A_{I}^{\text{FMA}}(\pi)&=\frac{\Delta}{2}\int_0^\pi {\exp}[ig\sin t]{\rm d}t
   \nonumber
   \\
   &= {\frac{\Delta\pi}{2}}[J_0(g)+iH_0(g)],
\end{align}
where \(H_0\) is the zeroth-order Struve function\cite{Abramowitz1964}. As depicted in Fig.~\ref{fig:eps_d}, the first-order symmetry conserving result is slightly more accurate than the one based only on \(J_0\), at the cost of a more complicated form for \(\epsilon\).

Furthermore, since \(\cos(\pi/2+t)=-\cos(\pi/2-t)\), \(H(t)\) and \(H_I(t)\) possess PT symmetry with respect to the reflection time \(t=\pi/2\) in the interval \(t\in[0, \pi]\). Therefore, as explained in Appendix~\ref{sec:symmetry_app}, the \(C_n(\pi)\) coefficients automatically vanish. That is, even-order ME terms for \(U(\pi)\) are all zero and thus the SMA does not improve on FMA. We stress, however, if \(\tilde{f}=\sin \omega t\) instead of  \(\tilde{f}=\cos \omega t\) is adopted, the symmetry is realized in 
another way.

In Fig.\ \ref{fig:eps_d} for $\Delta=\omega$, i.e., at the border of region {II}, we observe that especially for small values of $g$ there is still a discrepancy between the exact result and the lower order Magnus approximations. In contrast, TMA leads to a dramatic reduction in error, including the correction of the shifted relative extrema as well as the inaccurate positions of the first three exact quasi energy crossings observed in FMA. It is also noteworthy that all orders of the  symmetry conserving calculation show a nonzero slope at $g=0$, in agreement with the exact result, whereas this feature is absent from the \(J_0\) result . 
We note that for smaller values of $\Delta\ll 1$ 
inside region {II}, FMA is a very good approximation already (not shown).

\begin{figure}[htbp]
    \centering
\includegraphics[width=\linewidth]{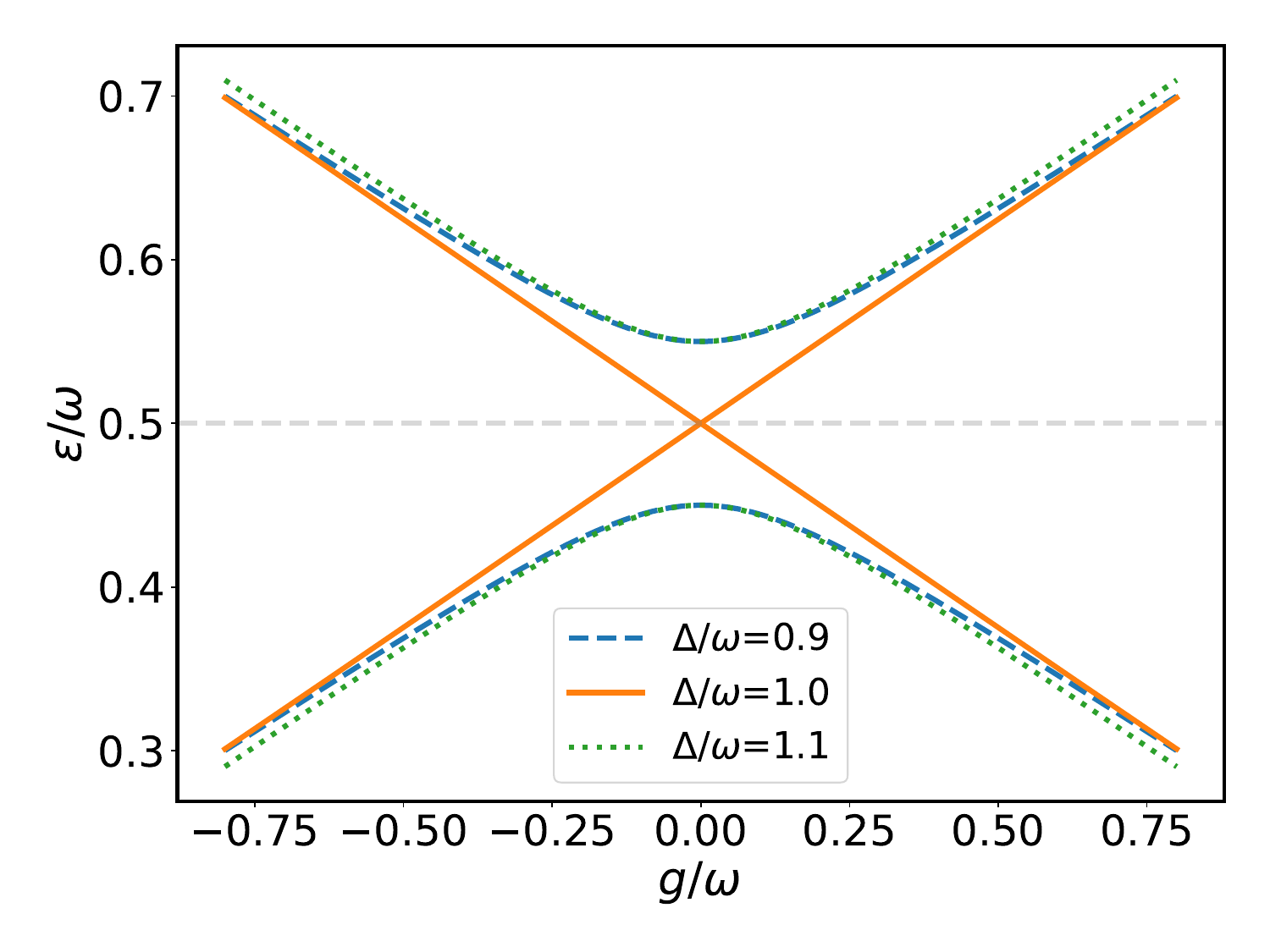}
    \caption{Exact results for $\epsilon$ in the Rabi model as a function of $g$ for different fixed \(\Delta\): when \(\Delta=\omega\) a coincident crossing takes place at \(g=0\), which results in non-zero slope (solid yellow line); for most values of \(\Delta\), the crossing is avoided and the slope is zero (\(\Delta=0.9\omega\): dashed blue, \(\Delta=1.1\omega\): dotted green).}
    \label{fig:eps_d_exact}
\end{figure}

\begin{figure}[htbp]
    \centering
    \includegraphics[width=\linewidth]{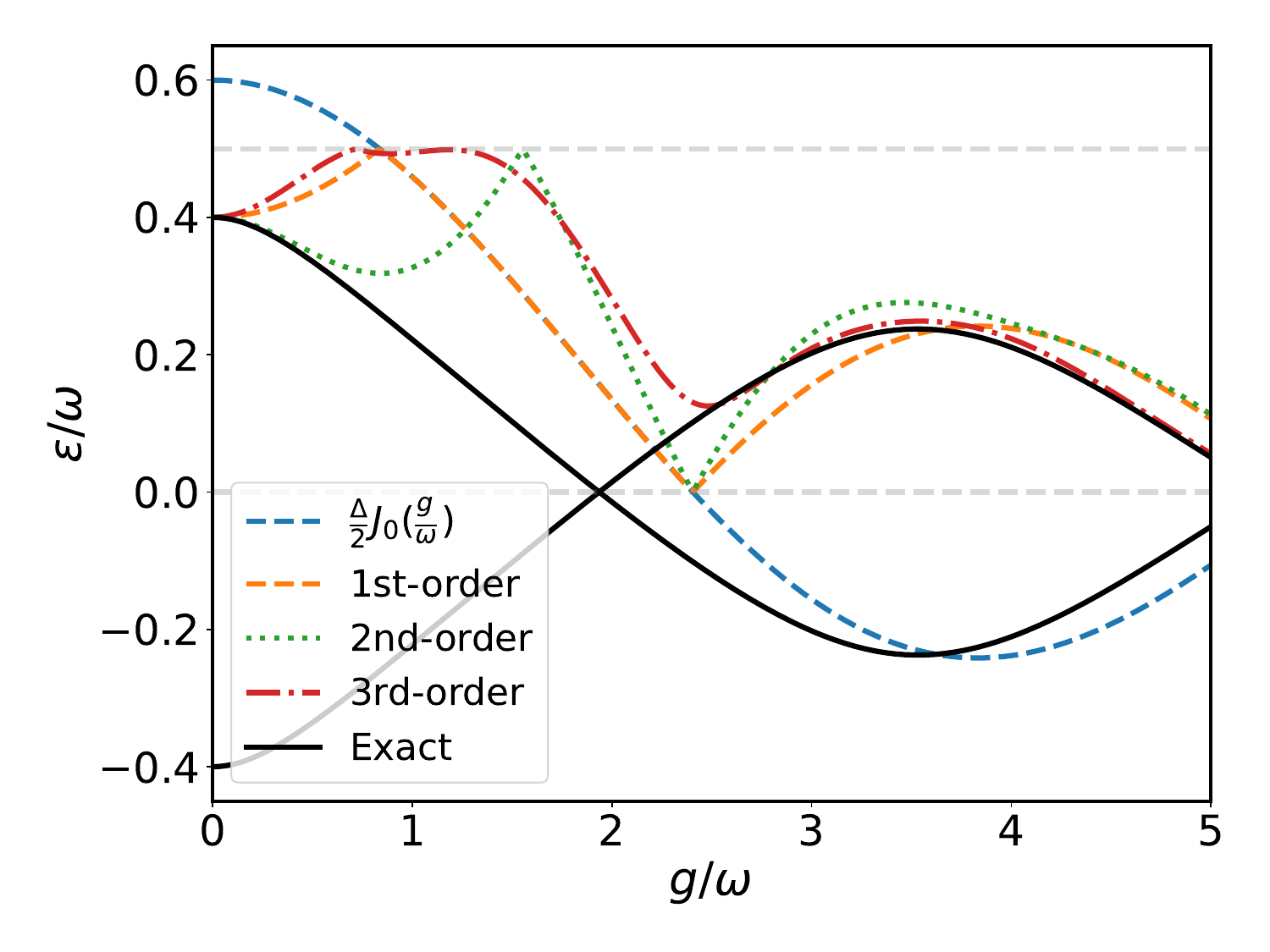}
    \caption{Different results for $\epsilon_{2\pi}$ in the Rabi model based on the picture appropriate for region II as a function of $g$ for \(\Delta/\omega=1.2\), outside of region II. Exact result for $\epsilon$ and $-\epsilon$ (solid black lines) and Magnus approximations of different order 
    (Bessel function results: dashed blue, 1st order: dashed orange, 2nd order: dotted green, 3rd order: dash-dotted red)
    for $\epsilon$, directly computed from \(U(2\pi)\)}
\label{fig:eps_d_comp}
\end{figure}

The non-zero slope at $g=0$ is related to a coincidental crossing on the boundary of 1st Brillouin zone, i.e. \(\epsilon=0.5\omega\). Actually, when $g=0$ the quasienergy obviously coincides with the static energy, namely,
\begin{equation}
    \epsilon_{g=0}= \pm \frac{\Delta}{2} \pmod{\omega}.
\end{equation}
For fixed \(\Delta\), two values (modulo \(\omega\)) of \(\epsilon\) must vary smoothly with \(g\) by virtue of the smooth dependence of solutions on parameters in linear systems. On the other hand, according to Eq.~\eqref{eq:Rabi_eps_symmetry}, in general \(\epsilon\) is an even function of \(g\) and thus has vanishing derivative at \(g=0\). The only exception arises when crossings take place, i.e., when \(\Delta/\omega\) is an integer. This is further illustrated in Fig.~\ref{fig:eps_d_exact} by the exact results of \(\epsilon\). Furthermore, we remark that the crossing at \(\epsilon=0.5\omega\) restricts the convergence radius of the ME\cite{Casas_2007}. When convergence of the ME is not guaranteed, spurious crossings may appear in the quasienergy spectrum, resulting in a complete loss of accuracy, as depicted in Fig.~\ref{fig:eps_d_comp}, where the different orders of the ME for $\epsilon_{2\pi}$ using the picture 
appropriate for region II are compared in the case $\Delta=1.2\omega$ (which is outside of region II!).

\subsubsection*{Region III}
Finally we address the adiabatic picture, which is expected to be a good approximation in region {  III}, when \(|g|>\omega, \ |\Delta| >\omega\). For this regime we choose the shape function as  \(\tilde{f}(t)=\cos(t)\) which is monotonic for \(t\in[0, \pi]\) so that that ME can be safely applied. As explained in Sec.~\ref{sec:non-adiabatic}, on the LZSM problem, the ME for \(U_a(\pi)\) actually converges for the entire \((g, \Delta)\) plane.

Firstly, in this case, the dynamical phase from Eq.\ \eqref{eq:dynamical_phase} reads
\begin{align}
\varphi(t)&=
{\frac{1}{2}}
\int_0^t\sqrt{\Delta^2+g^2\cos^2{r}}dr
\nonumber\\
&={\frac{1}{2}}
\int_0^t\sqrt{\Delta^2+g^2-g^2\sin^2{r}}dr=
\frac{m}{2}E(t,k),
\label{eq:incomp}
\end{align}
with \(m:=\sqrt{\Delta^2+g^2} , {k:=g^2/m^2}\) and where \(E\) is  Jacobi's {\it incomplete} elliptic integral of the second kind\cite{Abramowitz1964}. Secondly, we recall the transform into the adiabatic picture
from Eq.\ (\ref{eq:U_adiabatic_pic}), now taken 
at the half period:
\begin{equation}
U(\pi)=\Psi(\pi)\Phi(\pi)U_a(\pi)\Psi^\dagger(0).
\end{equation}
With the GP symmetry, one may verify that
\begin{equation}
P\Psi(\pi)P=\Psi(0)    
\end{equation}
and therefore
\begin{equation}
 PU(\pi)=    \Psi(0)\Phi(\pi)U_a(\pi)\Psi^\dagger(0)
=
\Phi(\pi)U_a(\pi).
\end{equation}
Here we omit the similarity transform, which does not change the trace in Eq.~\eqref{eq:eps_from_GPtr}. Using Eq.~\eqref{eq:BCH_b} and Eq.~\eqref{eq:eps_from_GPtr}, we 
can again extract the quasienergy from the time-evolution operator over half a period according to
\begin{equation}\label{eq:eps_adiabatic}
\epsilon_\pi =\frac{1}{\pi}
\sin^{-1}{\left\{\sin[\varphi(\pi)]\cos\theta_a
+
C_a\cos[\varphi(\pi)]
\frac{\sin\theta_a}{\theta_a}\right\}}.
\end{equation}

The plain adiabatic approximation 
corresponds to the zeroth order 
Magnus approximation (ZMA)
(\(A_a=C_a=\theta_a=0\)) and in
this limit, from Eq.~\eqref{eq:eps_adiabatic},
we get
\begin{equation}
    \epsilon_\pi^{\text{ZMA}} =\frac{1}{\pi} \varphi(\pi) \mod 1,
\end{equation}
which by using Eq.~\eqref{eq:incomp}, is  
given in terms of the {\it complete} 
elliptic integral of the second kind. The 
quasienergy thus is proportional to the 
dynamical  phase only, as can be seen by 
comparison with Eq.\ (\ref{eq:Berry}) and 
the above ZMA result is analogous to the 
one given by Shirley in his Eqn.\ 
(28)\cite{Shirley1965}, which was also 
derived for small frequencies.

\begin{figure}[htbp]
\centering
\includegraphics[width=\linewidth]{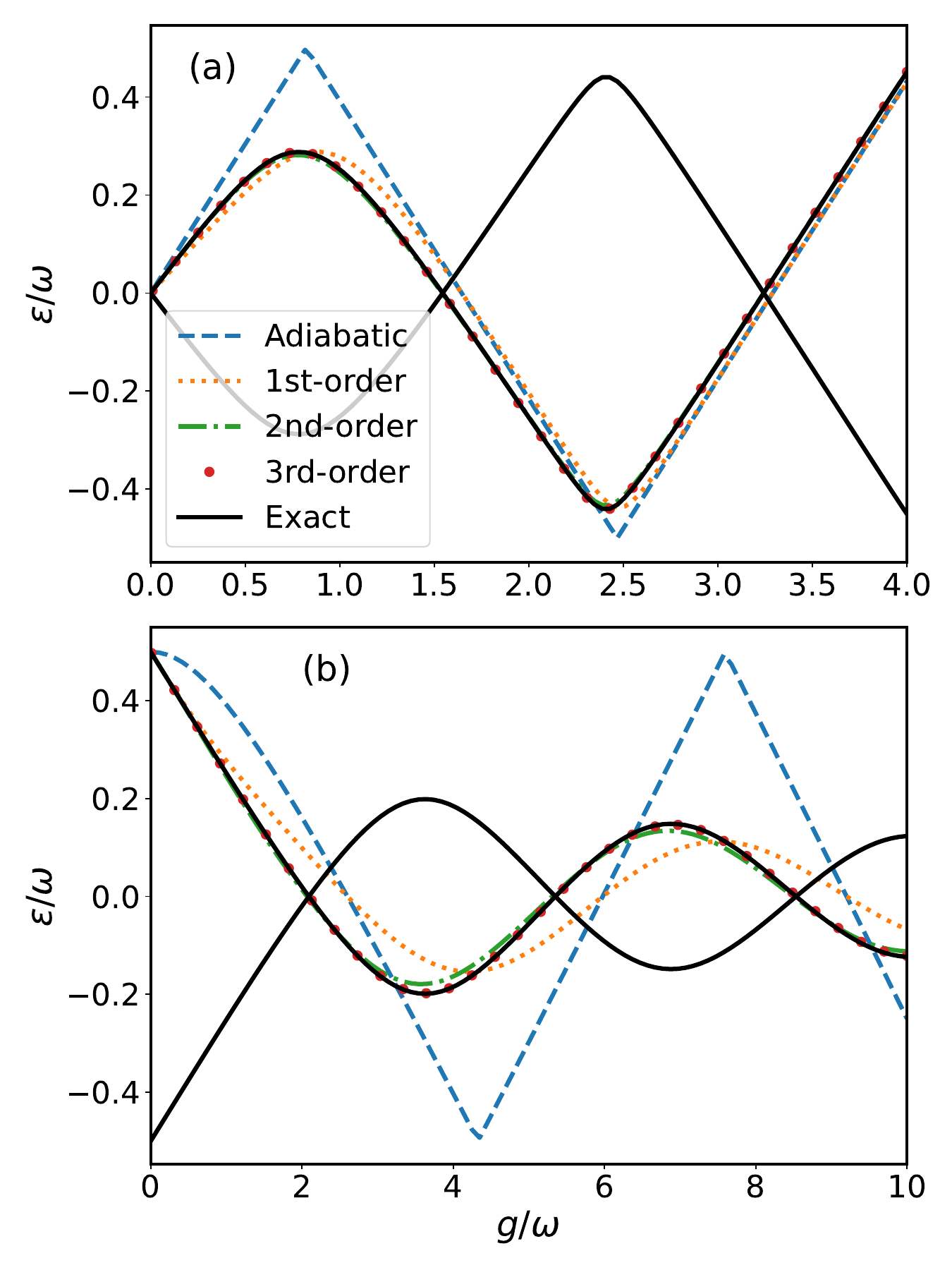}
\caption{Quasienergies of the Rabi model as a function of \(\Delta\) and \(g\): (a)  along the diagonal in the parameter plane displayed in Fig.\ \ref{fig:para}, i.e., for \(g=\Delta\): exact result for $\epsilon$ and $-\epsilon$ (solid black lines) and Magnus approximations of different order ( adiabatic: dashed blue, 1st order: dotted orange, 2nd order: dash-dotted green, 3rd order: red circles) in the adiabatic picture;}(b) same as (a) but with fixed \(\Delta/\omega=1\), i.e., we move along the horizontal line delineating regions III and II in Fig.\ \ref{fig:para}.
    \label{fig:eps_adiabatic}
\end{figure}
\begin{figure}[htbp]
\centering
\includegraphics[width=0.9\linewidth]{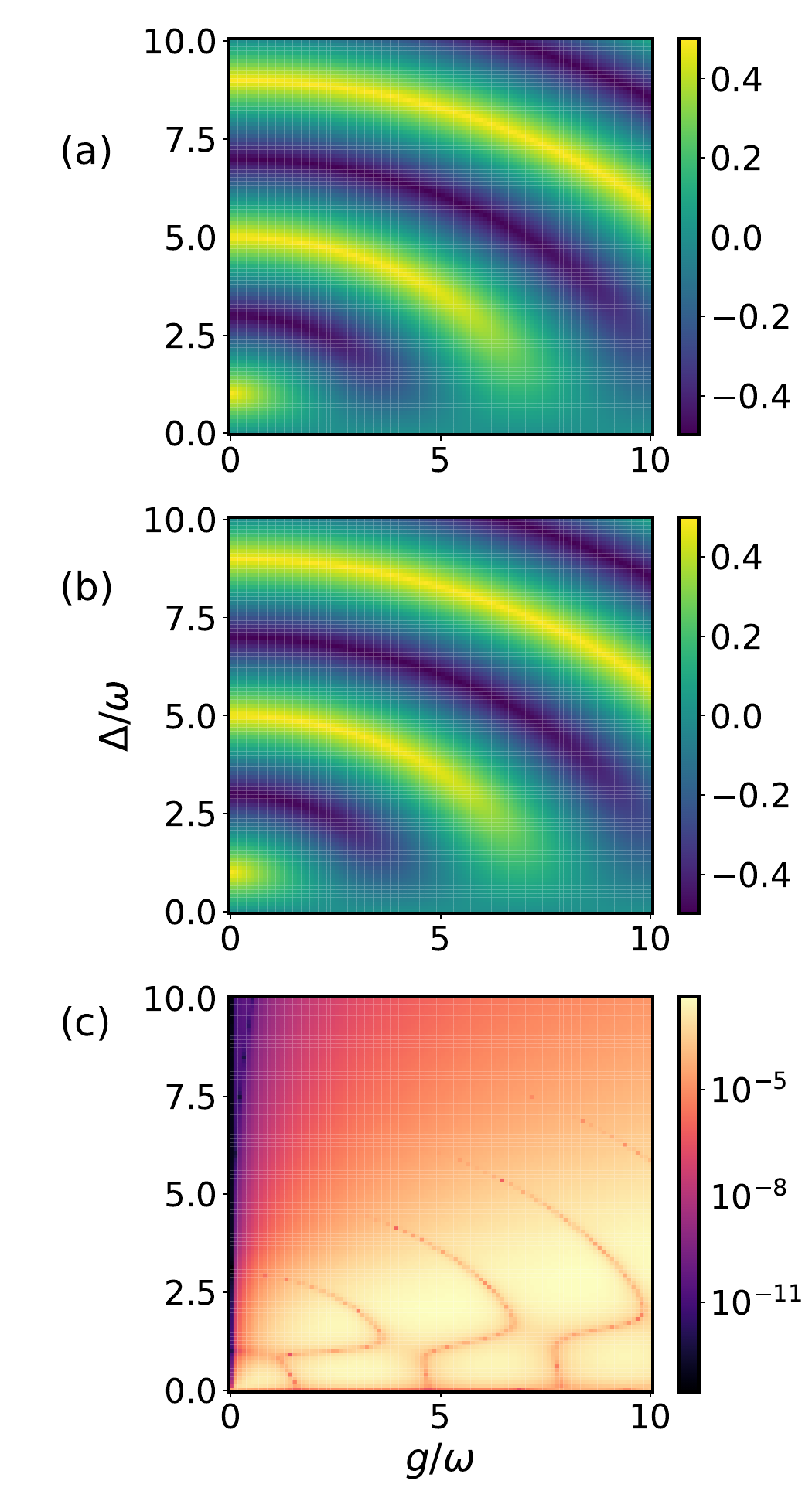}
\caption{Comparison between exact and third order Magnus results for quasienergies of the Rabi model as a function of \(\Delta\) and \(g\): (a) exact result; (b) third order Magnus approximation; (c) absolute error $|\epsilon_{\rm exact}-\epsilon_{\rm TMA}|$}
    \label{fig:eps_adiabatic_2d}
\end{figure}

Fig.~\ref{fig:eps_adiabatic} (a) shows the increasing accuracy of the  adiabatic result for large \(\Delta\) and \(g\), i.e., in the adiabatic limit of relatively small \(\omega\). Higher order Magnus approximations, on the other hand, also properly reveal the avoided crossings (between quasienergies from different Brillouin zones) at the extrema, with the most prominent ones at small values of $\Delta$(\(=g\)).  However, when \(\Delta/\omega\) is fixed, the adiabatic approximation breaks down for large external driving strength driving strength \(g\). In this case, as demonstrated in Fig.~\ref{fig:eps_adiabatic} (b), higher order Magnus approximations still may yield nearly exact results, thus vastly extending the applicability of the adiabatic approximation. To demonstrate the accuracy of TMA based on Eq.~\eqref{eq:eps_adiabatic}, we show the \(\epsilon(\Delta, g)\) in Fig.~\ref{fig:eps_adiabatic_2d}.
The extreme accuracy of the Magnus results can be seen there (the light yellow colors in panel (c) are already down to 10$^{-3}$).

\section{Conclusions and Outlook}\label{sec:discussion}

In this study we have demonstrated that the 
Magnus expansion may serve as a highly 
accurate approximation for the 
(semi-)analytical determination 
of non-perturbative transitions as well as 
of the quasienergy spectrum in (single-axis) 
driven two-level systems
already for relatively low orders if symmetry
properties are taken care of. 
For both models that we studied, the third 
order Magnus approximation 
was sufficient for almost complete 
agreement with the known exact analytical 
results in all of parameter space.

For the Landau-Zener problem, we 
have, e.g., shown that higher order 
Magnus approximations can not only 
accurately reproduce transition 
probabilities, but (from second order on) 
also yield the Stokes phase quite 
accurately. The PT symmetry of the model is 
preserved by the ME to 
all orders and the convergence criterion is 
always met. This is true for every monotonic 
drive and if this condition is not fulfilled, 
the drive can be split into monotony intervals 
and the Magnus approximations can be used for 
the separate domains.

It is well-known that the lowest order Magnus approximation for the semiclassical Rabi model has a similar accuracy as the rotating wave approximation\cite{Zeuch2020,DeyFloquetMagnus2025}.
By studying the time-evolution operator over
one period up to higher order Magnus approximation, we have
investigated the entire parameter regime of this model and have shown that exact and avoided crossings can be reproduced to a high degree. Especially, the Bloch-Siegert shift is quite accurately 
reproduced using a next-to-leading order 
Magnus approximation. The correct qualitative as
well as quantitative prediction of exact 
crossings calls for a ME based on the 
time-evolution operator over only half the period, however, 
which allows to restore the generalized parity 
symmetry that is lost otherwise. Results of third order are almost indistinguishable from the exact analytic results based on the confluent Heun function for the whole parameter space. The quality
of the results for the quasienergies is thus similar as that
of previous iterative approaches \cite{WaVa1998} but there is no restriction to the applicability of the ME if appropriate picture transformations are performed. If a picture is applied in a domain where it is not appropriate, the results are deteriorating rapidly as exemplified in the case of the Rabi model. An especially appealing picture is the adiabatic one, as has already been noted by Klarsfeld and Oteo, who found that the ME in the adiabatic pictures dramatically improves on perturbation theory for a spin 1/2 system in a rotating magnetic field \cite{Klarsfeld1992}. We have shown herein that, in the Rabi case and in the adiabatic picture, the second order ME result already gives almost perfect agreement with the exact analytic quasienergies in  a large parameter range.

On a technical note, we have emphasized the importance of the convergence condition, picture transformations and the maintenance 
of symmetry of the underlying problem. For monotonic driving, it has been shown that convergence criteria can be easily satisfied. However, in quantum control, the driving field is often modeled as a carrier oscillation modulated by a slowly varying envelope. It remains an important problem whether one can gain qualitative insights into the system’s behavior from the driving fields. A fully satisfactory answer remains elusive; nevertheless, our method appears promising and warrants further investigation. A very
interesting question, e.g., is if the application of the ideas herein improves on the findings by Begzjav and Eleuch on the two-level dynamics under an off-resonant pulse in the presence of a dissipative channel\cite{Begzjav2020103098}.

A simple yet effective twist of this study is to employ the \(\mathfrak{su}(2)\) algebra to decompose the Magnus expansion. This procedure can be generalized to the non-Hermitian case\cite{LDC25},  without significant changes. Furthermore, also multiple axis driving can and should be studied along the lines that we used herein. In addition, it seems to be a natural extension to apply the \(\mathfrak{su}(3)\) algebra\cite{Byrd-SU3} to treat three-level systems. Furthermore, the Magnus coefficients are very often expressed as multi-variate oscillating integrals. This indicates that that the Magnus approximation often becomes accurate under both large and small parameter limits, as we have noticed in the results of the LZSM and Rabi model. An error estimation analysis\cite{Lakos2023MagnusIIA, Blanes_1998} seems very appealing, yet it is far beyond the present study's scope.

Finally, in \cite{Petiziol2024} the ME is used to design an
effective driving along a third axis from a commutator of Pauli $\sigma_x$ and $\sigma_z$ matrices. To show this, the authors employed the Magnus-Floquet expansion to second order, whose accuracy depends on frequency. Yet our study shows that the validity of the ME is in general not limited by the frequency range, which might trigger new insights for the design of effective counter-diabatic Hamiltonians \cite{Berry2009} allowing adiabatic transitions within finite time\cite{Demirplak2003adiabatic}. In a similar vein, correction dissipators for a driven system have been derived in \cite{Balazs2025} by using  a truncated Magnus expansion. The ideas  presented here may help to optimize this
procedure.

\begin{acknowledgements}
{CW would like to acknowledge financial support by the ``Gesellschaft von Freunden und Förderern der TU Dresden e. V." in the form of a Deutschlandstipendium.}
\end{acknowledgements}

\vspace{0.5cm}
\noindent
{\bf AUTHOR DECLARATIONS}

\noindent
{\bf Conflict of interest}

The authors have no conflicts to disclose.

\noindent
{\bf Author contributions}

{\bf CW:} Conceptualization (lead), data curation (lead), formal analysis (lead), investigation (equal), methodology (lead), software (lead), validation (equal), visualization (lead),
writing-original draft (equal)
{\bf FG:} Conceptualization (supporting),
formal analysis (supporting), investigation (equal), supervision (lead), methodology (supporting), validation (equal), writing-original draft (equal)

\vspace{0.5cm}
\noindent
{\bf DATA AVAILABILITY}

The data that supports the findings of this study are available within the article. 

\appendix
\section{\(\mathfrak{su}(2)\) decomposition of the Magnus expansion}\label{sec:magnus_app}
Here we first review the Magnus expansion~\cite{Magnus1954} for a general Hamiltonian. Owing to the Schrödinger equation for the time-evolution operator
\(U(t)\) with initial condition \(U(t_0)=\mathbb{I}\), the Magnus operator \(\Omega(t)\) in Eq.~\eqref{eq:magnus_operator} must fulfill \cite{Magnus1954}
\begin{equation}\label{eq:Magnus_ODE}
     \dot\Omega(t)=\sum_{j=0}^{\infty}(-i)^j\frac{B_j}{j!}\text{ad}^j_\Omega(H(t)), \ \Omega(t_0)=0,
\end{equation}
where the adjoint operator is defined by \(\text{ad}_\Omega(\cdot)=[\Omega, \cdot]\), \(\text{ad}^k_\Omega(\cdot)=[\Omega, \text{ad}^{k-1}_\Omega(\cdot)]\) and \(B_0 = 1, B_1 = -1/2, B_2 = 1/6, B_3=0, B_4 = -1/30, ...\) are Bernoulli numbers.

Replacing the Hamiltonian \(H(t)\) 
by \(\lambda H(t)\) (and taking \(\lambda\) 
equal to \(1\) in the end), the Magnus 
expansion amounts to finding a power series 
expansion for the solution of 
Eq.~\eqref{eq:Magnus_ODE}, 
\begin{equation}
\label{eq:series}
\Omega(t;\lambda)=\sum_{n=1}^\infty\lambda^n\Omega_n(t).
\end{equation}
 In order to find \(\Omega_n\), one substitutes Eq.~\eqref{eq:series} into Eq. \eqref{eq:Magnus_ODE} and needs to collect all the \(\lambda^n\) contributions from both sides. 

To proceed, we are following the notation of Klarsfeld and Oteo \cite{MagnusRecursive1989}
and we define \(S_n^{j}(t)\) as the \(\lambda^n\) contribution from \((-1)^{j}\operatorname{ad}^j_\Omega(H(t))\). Eq.~\eqref{eq:Magnus_ODE} can be then written as
\begin{equation}
    \dot \Omega_1(t)=H(t), \dot \Omega_n(t)=\sum_{j=1}^{n-1}\frac{B_j}{j!}S_n^{(j)}(t) \ ( n\geq 2).
\end{equation}
Noting that
\begin{equation}
(-1)^{j}\operatorname{ad}^j_\Omega(H)=(-i)\bigl[\Omega, (-1)^{j-1}\operatorname{ad}^{j-1}_\Omega(H)\bigr],
\end{equation}
the quantity \( S_n^{j}\) can be thus recursively determined by
\begin{equation}\label{eq:Magnus_recursion}
    S_n^{(j)}(t)=\sum_{m=1}^{n-j}(-i)[\Omega_m(t), S_{n-m}^{(j-1)}(t)], \ 1 \leq j\leq n-1,
\end{equation}
with initial terms \(S_1^{(0)}(t)\equiv H(t), S_n^{(0)}(t):=0 \ (n\geq2)\). For reference, the first three orders of the Magnus operator read\cite{Tannor2007}
\begin{align}
\Omega_1(t) &= \int_{t_0}^t {\rm d}t_1\, H(t_1), \\[6pt]
\Omega_2(t) &= -\frac{i}{2}\int_{t_0}^t {\rm d}t_1 \int_{t_0}^{t_1} {\rm d}t_2\; [\,H(t_1),\,H(t_2)\,], \\[6pt]
\Omega_3(t) &= -\frac{1}{6}\int_{t_0}^t {\rm d}t_1 \int_{t_0}^{t_1} {\rm d}t_2 \int_{t_0}^{t_2} {\rm d}t_3\;
\nonumber 
\\ &\quad
\bigl([\,H(t_1),[\,H(t_2),H(t_3)\,]\,] 
\nonumber 
\\
&+ [\,H(t_3),[\,H(t_2),H(t_1)\,]\,]\bigr). 
\end{align}

Now we specialize the recursion relation to a  Hamiltonian in \(\mathfrak{su}(2)\). Using the standard commutation relations
\begin{equation}
    \left[\vec{n}_A\cdot \vec{\sigma}, \vec{n}_B\cdot \vec{\sigma}\right]=2i(\vec{n}_A\times \vec{n}_B)\cdot \vec{\sigma},\label{eq:su2_commutator}
\end{equation}from Eq.~\eqref{eq:Magnus_recursion} it follows 
that \(S_n^{(j)}(t)\)  is also an element of  \(\mathfrak{su}(2)\) at any time, and thus it can be decomposed as
\begin{equation}\label{eq:decomp_S_nj}
    S_n^{(j)}=a_n^{(j)}(t)\sigma_+ +a_n^{(j)*}(t)\sigma_- + c_n^{(j)}(t)\sigma_z.
\end{equation}
Then the decomposition of \(\Omega_n(t)\) follows to  be
\begin{equation}\label{eq:decomp_Omega_n}
\Omega_n(t)= A_n(t)\sigma_+ +A^*_n(t)\sigma_- + C_n(t)\sigma_z,
\end{equation}
where
\begin{align}
    A_n(t)=&\sum_{j=0}^{n-1}\frac{B_j}{j!}\int_{t_0}^t a_n^{(j)}(\tau){\rm d}\tau, 
\\ 
C_n(t)=&\sum_{j=0}^{n-1}\frac{B_j}{j!}\int_{t_0}^t c_n^{(j)}(\tau){\rm d}\tau.
\end{align}

By substituting Eqs.~\eqref{eq:decomp_S_nj}-\eqref{eq:decomp_Omega_n} into Eq.~\eqref{eq:Magnus_recursion} and using 
the \(\mathfrak{su}(2)\)  commutation relations in Eq.~\eqref{eq:su2_algebra}, the recursion relation can be reduced into commutator-free form:
\begin{align}
    a_n^{(j)}(t)=&2i\sum_{m=1}^{n-j}\left[
A_m(t)c_{n-m}^{(j-1)}(t) - C_m(t)a_{n-m}^{(j-1)}(t) \label{eq:recursion_a}
\right],
\\
c_n^{(j)}(t)=&2\sum_{m=1}^{n-j}\text{Im}\left[A_m(t) a_{n-m}^{(j-1)*}(t)\right],\label{eq:recursion_c}
\\
a_{n}^{(0)}(t)=&c_{n}^{(0)}(t)=0\quad  (2\leq n,\  1 \leq j\leq n-1).\label{eq:recursion_boudary}
\end{align}
Recalling that \(S_1^{(0)}(t)\equiv H(t)\), the initial  terms for the above recursion are determined by the \(\mathfrak{su}(2)\) decomposition 
\begin{equation}\label{eq:ac_initial}
    H(t) = a_1^{(0)}(t)\sigma^+ +a_1^{(0)*}(t)\sigma^-+c_1^{(0)}(t)\sigma_z
\end{equation}
of the Hamiltonian.

Finally, we emphasize that the recursion relations, Eqs.~\eqref{eq:recursion_a}–\eqref{eq:recursion_boudary}, are particularly amenable to algorithmic implementation, as pointed out in~\cite{MagnusRecursive1989}. At each order, no more than two accumulative integrals are required, which allows for efficient numerical evaluation of the coefficients, at least to leading orders.
\section{Parity-time symmetry of the Magnus operator}\label{sec:symmetry_app}

In this appendix we discuss how the parity-time (PT) symmetry is carried over to the Magnus operator level. For the reader's convenience, we
repeat the definition of PT symmetry 
\begin{equation}\label{eq:PT_symmetry}
    PH^*(-t)P=H(t),
\end{equation}
from Eq.\ \eqref{eq:PT}, where the parity operator is Hermitian as well as unitary, i.e., it satisfies
\begin{equation}
    P^\dagger = P, \quad P^2=\mathbb{I}.
\end{equation}
In addition in this appendix we  propose 
that \(P\) has a real (matrix) representation, 
i.e.,
\begin{equation}\label{eq:P_real}
    P^*=P
\end{equation}
is assumed to hold.

On the time-evolution operator level, according to the Schrödinger equation, one may directly verify that PT symmetry leads to the following result:
\begin{equation}\label{eq:PT_on_U}
    U^*(-t, 0)=PU(t,0)P.
\end{equation}
 We now consider the Magnus operator on a symmetric time interval \([-t_f, t_f]\). 
For simplicity, we denote
\begin{equation}
\label{eq:defS}
S\equiv U(t_f, -t_f)\equiv \exp(-i\Omega).  
\end{equation}
By the group property of the evolution operator, we have
    \begin{align}
        S&=U(t_f, 0)U(0,-t_f)\notag\\
        &=U(t_f,0)U^{-1}(-t_f, 0)\notag\\
        &=U(t_f,0)P[U^*(t_f,0)]^{-1}P.
    \end{align}
In the last step we have used Eq.~\eqref{eq:PT_on_U} and the properties of \(P\). It thus follows that 
\begin{equation}
    [S^*]^{-1}=PSP.
\end{equation}
On the other hand, according to Eq.~\eqref{eq:defS} and the definition of matrix exponent we have
\begin{equation}
    [S^*]^{-1}=\exp (-i\Omega^*)
\end{equation}
and 
\begin{equation}
    PSP=\exp[-i(P\Omega P)].
\end{equation}
Equating the two last equations, finally, we conclude that
\begin{equation}\label{eq:PT_on_Omega}
    \Omega^*=P\Omega P.
\end{equation}
Note that \(\Omega\) as well as \(\Omega^\ast\) are both associated with the time interval \([-t_f,t_f]\). 

Remarkably, since the Magnus expansion is essentially a power series expansion of \(\Omega\), as given in Eq.\ (\ref{eq:series}), the property Eq.~\eqref{eq:PT_on_Omega} holds to each order. We note in passing that to reach 
the final conclusion, the Hamiltonian does not have to be Hermitian.

To illustrate the application of Eq.~\eqref{eq:PT_on_Omega}, we elaborate on the PT symmetry of the two-level Hermitian systems from 
the main text. For this, we rewrite Eq.~\eqref{eq:Omega_decomp} as
\begin{equation}\label{eq:Omega_xyz}
    \Omega =\text{Re}(A)\sigma_x-\text{Im}(A)\sigma_y+C\sigma_z,
\end{equation}
and thus (note that $C$ is real-valued)
we find
\begin{equation}
    \Omega^* =\text{Re}(A)\sigma_x+\text{Im}(A)\sigma_y+C\sigma_z.
\end{equation}
Due to the symmetry expressed by Eq.~\eqref{eq:PT_on_Omega}, certain  coefficient in Eq.~\eqref{eq:Omega_xyz} may vanish, depending on the choice of \(P\):
\begin{enumerate}
    \item When \(P=\sigma_z\), we have
\begin{equation}
    P\Omega P=-\text{Re}(A)\sigma_x+\text{Im}(A)\sigma_y+C\sigma_z,
\end{equation}
and thus \(\text{Re}(A)=0\), as we have seen in the discussion of LZSM model in Sec.~\ref{sec:non-adiabatic}.
    \item When \(P=\sigma_x\), we have
\begin{equation}
    P\Omega P=\text{Re}(A)\sigma_x+\text{Im}(A)\sigma_y-C\sigma_z,
\end{equation}
leading to \(C=0\), as in region II for the Rabi model, discussed in Sec.~\ref{sec:Floquet}.
    \item When \(f(-t)=f(t)\in \mathbb{R}\) in Eq.~\eqref{eq:phys_hamiltonian}, the Hamiltonian displays time reflection symmetry, i.e., \(H^*(-t)=H(t)\). Formally, this amounts to taking \(P=\sigma_0\) in Eq.~\eqref{eq:PT_symmetry}, which results in \(\text{Im}(A)=0\).
    {This symmetry, e.g., holds for \(t\in[0, \pi/\omega]\), reflected about \(t=\pi/2\omega\), if the shape function \(\tilde{f}(t)=\sin\omega t\) is used in the Rabi model.}
\end{enumerate}
\section{Exact and avoided crossings of quasienergy in two-level Floquet systems}\label{sec:crossing_app}
Here we discuss the quasienergy crossings in a periodically driven two-level system in greater detail. Due to the multi-valuedness
of $\epsilon$ (which is defined modulo $\omega=1$), quasienergy crossings
can be grouped into two classes:
\begin{enumerate}[label=\arabic*), ref=\arabic*)]
    \item \label{itm:crossing1} crossing in the middle of the Brillouin zone:
\[\epsilon =-\epsilon \Rightarrow\epsilon=0\Leftrightarrow U(2\pi)=\sigma_0\]
    \item \label{itm:crossing2} crossing of different branches (at the boundary of the Brillouin zone):
\[\epsilon =-\epsilon\pm 1 \Rightarrow\epsilon=\pm\frac{1}{2}\Leftrightarrow U(2\pi)=-\sigma_0\]
\end{enumerate}
In both cases, any initial state reproduces itself after one period, up to a global phase factor. This provides a two-level interpretation of the coherent destruction of tunneling \cite{CDT-PRL-1991}, originally discovered in the periodically driven symmetric double-well potential model. In general, because an element in SU(2) is characterized by three free parameters, realizing the crossing requires at least tuning three coupling constants in the Hamiltonian. This constitutes a particular case of the Wigner-von Neumann non-crossing rule \cite{vonNeumann1929}. Exact crossings become much easier to achieve in the presence of symmetry, however, as explained below. 

When \(\tilde{f}(t)\) has the property
\(\tilde{f}(t+\pi)=-\tilde{f}(t)\), the Hamiltonian Eq.~\eqref{eq:phys_hamiltonian} displays the following adjoint symmetry
\begin{equation}\label{eq:generalized_parity}
    PH(t+\pi)P=H(t),
\end{equation}
where \(P=\sigma_z\) in this case. More generally, we define Eq.~\eqref{eq:generalized_parity} as the generalized parity (GP)
by allowing \(P\) to be a unitary Hermitian operator, satisfying
\begin{equation}
P^\dagger=P,\quad P^2=\sigma_0, \quad \det{P}=-1.
\end{equation}
And thus \(P\) can be represented as
\begin{equation}
P=\vec{n}_P \cdot \vec{\sigma}, \quad |\vec{n}_P|=1.
\end{equation}
With Eq.~\eqref{eq:generalized_parity} one may prove\cite{Schmidt2019} that
\begin{equation}\label{eq:U_2pi_GP}
U(2\pi)=\left[PU(\pi)\right]^2.
\end{equation}
It follows from Eq.~\eqref{eq:U_2pi_GP} that \(PU(\pi)\) has eigenvalues \(\mp e^{\pm i\epsilon \pi}\) where \(\epsilon\) is the quasienergy and its eigenstates are the initial Floquet states. Thus we have
\begin{equation}
    \text{tr}[PU(\pi)]=-2i\sin{(\epsilon \pi)}.
\end{equation}
On the other hand, by writing \(U(\pi)\) in angle-axis form \(\exp(-i\theta \vec{n}\cdot \vec{\sigma})\), one finds
\begin{equation}\label{eq:tr_u_GP}
{ \text{tr}[PU(\pi)]=-2i(\vec{n}\cdot\vec{n}_P)\sin{\theta}}.
\end{equation}
Comparing the two equations above, one obtains
\begin{equation}\label{eq:eps_from_GPtr_app}
    \sin{(\epsilon \pi)}=(\vec{n}\cdot\vec{n}_P)\sin{\theta}.
\end{equation}
Eq.~\eqref{eq:eps_from_GPtr_app} explains explicitly the crossing and non-crossing rules of the quasienergies by the following argument. For case \ref{itm:crossing1} to appear,
\begin{equation}
 \sin{\epsilon\pi} =0\Leftrightarrow \theta =0 \text{ \it or } \vec{n}\cdot \vec{n}_P=0.
\end{equation}
The set of allowed \(U(\pi)\) thus is two-dimensional, and therefore it suffices to tune only one parameter to achieve an exact crossing at \(\epsilon=0\) in general. In contrast, case \ref{itm:crossing2} is equivalent with
\begin{equation}
\sin{\epsilon\pi}=\pm 1  \Leftrightarrow \theta=\pm \frac{\pi}{2} \text{ \it and } \vec{n}=\pm \vec{n}_P,
\end{equation}
that is, \(\epsilon=\pm 1/2\) and \(U(\pi)=\pm i\sqrt{P}\). This case is much harder to realize since the allowed set of \(U(\pi)\) has zero measure in the SU(2) group. Therefore under change of parameters, around \(\epsilon=\pm 1/2\), i.e. at boundaries of the Brillouin zone, crossings are usually avoided. 



\nocite{*}
\bibliography{reference}

\end{document}